\documentclass[apj]{emulateapj}
\setkeys{Gin}{width=0.45\textwidth}
\usepackage{graphicx}
\usepackage{amsmath}

\newcommand{\lsun}{{\rm\,L_\odot}}
\newcommand{\hmpc}{\ifmmode{h^{-1}\,\hbox{Mpc}}\else{$h^{-1}$\thinspace Mpc}\fi}

\newcommand{\kms}{\ifmmode{\,\hbox{km\,s}^{-1}}\else {\rm\,km\,s$^{-1}$}\fi}

\begin{document}
\title{Star Streams in Triaxial Isochrone Potentials with Sub-Halos}
\shorttitle{Star Stream Velocity Substructure }
\shortauthors{Carlberg}
\author{R. G. Carlberg\altaffilmark{1,2} }
\altaffiltext{1}{Department of Astronomy and Astrophysics, University of Toronto, Toronto, ON M5S~3H4, Canada} 
\altaffiltext{2}{Canadian Institute for Advanced Research, Cosmology and Gravity Program, Toronto, ON M5G 1Z8}
\email{carlberg@astro.utoronto.ca }

\begin{abstract}
The velocity, position, and action variable evolution of a tidal stream drawn out of a star cluster 
in a triaxial isochrone potential containing a sub-halo population reproduces many of the orbital effects of more general
cosmological halos but allows easy calculation of orbital actions.
We employ a spherical shell code which we show 
accurately reproduces the results of a tree gravity code for a collisionless
star cluster.
Streams from clusters on high eccentricity orbits, $e\gtrsim 0.6$, can spread out so much that the
amount of material at high enough surface density to stand out on the sky may be only a few
percent of the stream's total mass.
Low eccentricity streams remain more spatially coherent, but sub-halos both broaden the stream 
and displace the centerline with details depending on the orbits allowed within the potential. 
Overall,
the majority of stream particles have changes in their total actions of only 1-2\%, leaving 
the mean stream relatively undisturbed.
A halo with 1\% of the mass in sub-halos typically spreads the velocity distribution about a factor of 
two wider than would be expected for a smooth halo. Strong density variations, ``gaps'', along with
mean velocity offsets, are clearly detected in low
eccentricity streams for even a 0.2\% sub-halo mass fraction.
Around one hundred velocity measurements per kiloparsec of stream will enable tests for 
the presence of a local sub-halo density as
small as 0.2-0.5\% of the local mass density, with about 1\% predicted for 30 kiloparsec orbital radii streams.
\end{abstract}

\section{INTRODUCTION}
\nobreak

Tidal star streams pulled out from star clusters have been extensively studied in simulations where the galaxy is modeled
as an analytic potential, often taken to be spherically or axially symmetric for simplicity
\citep{McGlynn:90,Johnston:96,Dehnen:04,Kupper:08,Varghese:11, Kupper:12,Amorisco:15,Fardal:14,
NC:14, Carlberg:15, Hozumi:15}. 
These simulations have described how tidal heating ejects stars into a stream.
There is a good analytic understanding of how sub-halos 
interact with thin streams in angular momentum conserving potentials
 to create velocity perturbations which open up to gaps in a stream \citep{Yoon:11, Carlberg:12, Erkal:15}.
Simulations that
begin with a cosmological halo show considerably more orbit diffusion than 
smooth analytic potentials \citep{Bonaca:14, Pearson:15, Ngan:15a, Ngan:15b} which is not yet well understood.
However, the increased orbital freedom is certainly related to 
the rich orbit structure in static triaxial potentials \citep{deZeeuw:85,BT:08} 
which is bound to play a role in the aspherical and non-stationary cosmological galactic halos in 
which tidal streams develop.

In principle, kinematic measurements along a stream can tightly constrain the full phase space parameters 
of the highly redundant orbits of the stream particles and can serve as a precision probe of a galactic potential 
\citep{Johnston:99,Binney:08,Bovy:14}.
The initial velocities of streams of stars that are tidally stripped from star clusters  are
calculated for a given orbit and potential with collisionless n-body codes. 
The stars emerge with highly correlated orbital velocity substructure \citep{Carlberg:15}.
The stream width scales with the tidal radius and the length of 
 scales with the tidal radius and the age of the stream \citep{Johnston:98}.
In a spherical potential 
angular momentum and radial action are always conserved, but tidal
mass loss features blur out with orbital overlap down the stream. 
However
as stars move away from the progenitor on slightly different orbits in which the gravitational field has variations
due to aspherical structure in the halo the streams can disperse relative to to a simple smooth potential model
\citep{Bonaca:14, Ngan:15b}. Relative to a spherical potential the streams are almost everywhere increased
in width, with some portions greatly blurred. However, a few segments of  high or even enhanced
density can remain \citep{Ngan:15b}. 
The view of the stream in the dynamical action variables, which are adiabatically conserved
quantities of a potential, can provide insight into what is causing the stream to change its appearance \citep{Ibata:02}.

The purpose of this paper is to quantitatively analyze
the dynamical behavior of tidal streams in moderately aspherical potentials 
that also contain a sub-halo population.  
The analysis will be done
within a simplified potential which is able to reproduce many of the features of a cosmological halo but allows the
calculation of action-angle variables at the beginning and end of the simulation.
A simplified n-body code that reproduces the tidal stream from a standard n-body code is used to speed up the simulations.
The changes in angular momentum and the radial action along 
a stream 
as a function of the halo shape and sub-halo content  appear as  changes to the stream in position and velocity space. 
These results will be useful to advance the understanding of the mass structure and contents of galactic
halos as more and better kinematic data become available and serve as a guide to more realistic simulations.


\section{Orbits in Triaxial Isochrone Potentials}

We want to analyze the orbit changes of a stellar stream in a galactic halo with sub-halos.  
Action variables, as conserved quantities, provide direct insight 
into what has changed. 
Henon's isochrone potential  has simple algebraic expression for the actions \citep{Henon:I,Henon:II} and
provides a reasonable match to a galactic rotation curve over the range of radii where we want to study stream dynamics.  
One fairly standard way to create a triaxial potential is to apply a linear scaling to the Cartesian coordinates
of a spherical potential to create a triaxial potential.
The simplicity of the axis scaling comes at the cost of losing the analytic properties of the orbits in the
spherical potential. 
A number of axisymmetric potentials with good analytic properties have been developed
\citep{EB:11,Binney:14} but these still do not have the triaxiality, or more generally, the asphericity, of cosmological halos. 

With the goal of allowing the rich orbital structure of a triaxial potential yet retaining some of the analytic
properties of the isochrone, we introduce a time dependent potential, 
 which for  cosmological halos should be a reasonable approximation.
The potential starts as a spherical isochrone, makes a smooth transition to a triaxial form, then
remains as a triaxial for a relatively long interval, then reverses to become spherical again. 
The scale radius and the mass remain constant, the shape being only thing that changes. We will 
leave the $x$ axis unaltered, but compress the $y-z$ axes, which alters the density profile, increasing 
the density in the central region and generally leading to negative densities at large radius on the most compressed axis.
We will also smoothly turn on and off a small fraction of the mass in an identically distributed sub-halo population.

The potential's spherical initial and final form is a device
to allow the analytic actions to be calculated and is not intended to be realistic.
The action variables of the spherical isochrone will be valid quantities to calculate
at the beginning and end of the simulation, although they are not conserved quantities over the
simulation. 
We expect that there
will be smooth changes in the action variables associated with the turn-on and turn-off of the triaxiality, and
local changes as a result of sub-halo encounters, the two being sufficiently different that it might
be possible to untangle the effects. 
Based on these considerations the triaxial isochrone potential used here is,
\begin{equation}
\Phi(x,y,z,t)={-GM(t)\over{b+\sqrt{b^2 +x^2+y^2q_y^{-2}(t)+z^2q_z^{-2}(t)}}}.
\label{eq_isot}
\end{equation} 
The $q_y(t)$ and $q_z(t)$ are dimensionless functions which begin and end at unity, 
at which time the potential  is spherically symmetric. 
We largely focus on compressions where the $q$ values are less than one,
 but cosmological halos also undergo small expansions during mergers with other halos as they.
Our standard compression has  $q_y=0.9$ and $q_z=0.95$ which leads to small negative 
densities on the $y$ axis beyond $y=6.51$, with a negative peak at $y=8.83$ with about $10^{-4}$ of the central density. 
The negative densities occur outside of the orbits that will be studied.
The gravitational force is always inward directed.

\subsection{Time Variation of Potential Shape}

The potential asphericity turn-on takes place over a time $t_r$, after which it remains on for a time $t_f$, using
the arbritrary function,
\begin{align}
A(t) = & 2(t/t_r)^2  & 0\le t \le t_r/2,  \nonumber\\
	= & 1-2(1-t/t_r)^2 & t_r/2< t\le t_r,  \nonumber \\
	=& 1 & t_r<t <t_f+t_r. 
\label{eq_At}
\end{align}
For $t>t_f+t_r$ the rising function is reversed and $A(t)$ returns smoothly back to zero.
The rise time, $t_r$, is chosen to  reasonably slow, typically about 3 radial orbits, which is neither in the impulsive
regime nor truly in the adiabatic regime. The time 
spent as a constant triaxial potential, $t_f$, is typically about 15 radial orbits, 
comparable to the time over
which thin streams have been developing in the Milky Way halo. 

All calculations here use units with $GM=1$ and $b=1$. 
We usually start the calculation with a cluster placed
at  an $x$ position of 4 units. We characterize the
starting conditions with the angular momentum, $L$, relative to 
the angular momentum of a circular orbit in the spherical potential, $L_c$.
Orbits with
$L/L_c$ of 0.4 and 0.7 have radial orbital periods are 29.2 and 35.3, respectively, in the spherical
potential when started at $r=4$. For the time variation function $A(t)$ we
select $t_r=80$ and $t_f=440$,  so that the potential becomes spherical again at time 600 and the simulation terminates.

The co-ordinate distortions are evaluated as
 $q_y(t)=1+a_yA(t)$ and $q_z(t)=1+a_zA(t)$. 
Positive $a_y, a_z$ correspond to an expansion of the potential, that is,
 reduced gravitational attraction at locations off the $x$ axis and 
negative values correspond to a compression. 
Both expansion and compression are possible in cosmological situations, although compression 
likely dominates and is relatively small, since
accreted mass spends little time in the central region and largely deposits mass near the current virial radius of the halo.
Cosmological halos also can have a rotation of the potential shape, which we will ignore at this time.
\citet{Louis:88} studied radial perturbations to the isochrone potential finding that higher order resonances quickly 
appeared along with complex orbits, leading us to expect behavior of that sort.

\subsection{Orbiting Sub-Halos}

Sub-halos are present in large numbers in the dark matter halos created in cosmological simulations
\citep{Klypin:99,Moore:99,VL1,Aquarius,Stadel:09}. 
The differential mass distribution is $m\,dN/dm \propto m^{-0.9}$ hence has nearly a constant
mass per decade of mass, $m^2\,dN/dm \propto m^{0.1}$.  
The mass fraction 
of sub-halos varies steeply with radius relative to the local dark matter density in cosmological halos  \citep{Aquarius}. 
For the radial range around 30 kpc where the few known thin streams have been found the models predict about
1\% of the mass is in sub-halos, with the sub-halo mass fraction dropping to about 0.1\% at a nominal
solar radius of 8 kpc.

For the interests of this paper the practical issue is to have a well
defined sub-halo population over the radial range of the star cluster orbit.  
Conceptually our approach is to convert a small fraction,  $f_{sh}$, of the mass of the fixed isochrone 
potential into a set of sub-halos with the same collective massive distribution as the isochrone. 
The sub-halos are free to orbit
within the isochrone potential as test particles, which requires that they be randomly drawn from a 
distribution function which generates
the isochrone potential. Conveniently analytic forms are available for the isochrone's mass and energy 
distribution functions \citep{Henon:III,Binney:14}.  
Therefore  on the average there is no change in the summed sub-halo plus diminished external potential, and, 
on the average, the global and local sub-halo
fractions are equal everywhere for the initial spherical potential.
Although the density distribution of the isochrone extends to infinity, 59\% of the mass is within the
$r=4$ orbit
where we start the star clusters.
Hence, little computational effort is wasted on the relatively few sub-clusters at larger radii which do not have orbital pericenters 
within the range of radii that the star cluster explores. 
The sub-halos respond to the undiminished triaxial isochrone 
potential but not each other. 
The stars of the cluster and the tidal streams feel gravitational forces from all other stars, plus the isochrone
reduced by $(1-f_{sh}A(t))$ plus the sum of the forces from all the sub-halos which vary as $f_{sh}A(t)$.
where the $A(t)$ factor, Equation~\ref{eq_At}, which rises from zero 
and returns to zero at the end of the simulation.

All the sub-halos are given identical masses, thereby allowing us to examine mass dependent effects.
That is, once a number, $N_{sh}$, of sub-halos is chosen, the sub-halo all have mass, $m_{sh}=f_{sh}/N_{sh}$.
 Each sub-halo is modeled as a Herquist potential,$-Gm/(r+a)$, \citep{Hernquist:90}, with a scale radius 
$a_{sh}=\sqrt[3]{m_{sh}}$,  
as expected for cosmological sub-halos of constant 
scale density. The dependence of the results on the 
scale radius of the sub-halos are visible in the simplified analytic analysis \citep{Carlberg:15,Erkal:15}, where there is little change
beyond the scale radius, but the induced velocities increase for stars that have encounters that take them within
the scale radius of a sub-halo.

\begin{figure}
\begin{center}
\includegraphics[angle=-90, scale=0.8]{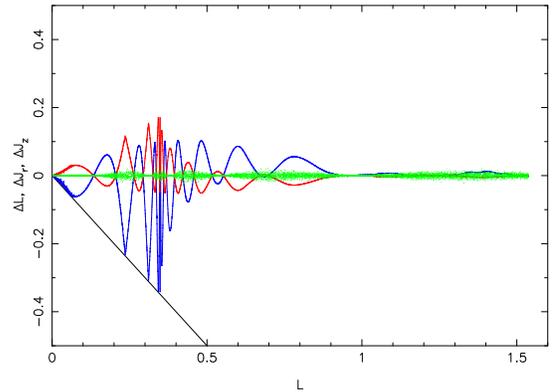} 
\end{center}
\caption{The changes between beginning and end of the simulation in the particle's total angular momentum, 
$L$ (blue), radial action $J_r$ (red) and angular momentum
about the axis perpendicular to the initial orbital plane, $J_z$ (green),
as a function of the initial $L$ in a potential with  $a_y=-0.05$ and $a_z=-0.10$.
The particles are started at $r=4$ with a uniform angular momentum distribution
between 0 and the circular value. All orbits are started in the plane
with an initial vertical and radial velocity dispersion of 0.002 units.  A point below the downward sloping
line drawn from the origin would have had its orbital direction reversed.}
\label{fig_iso145}
\end{figure}

\begin{figure}
\begin{center}
\includegraphics[angle=-90, scale=0.8]{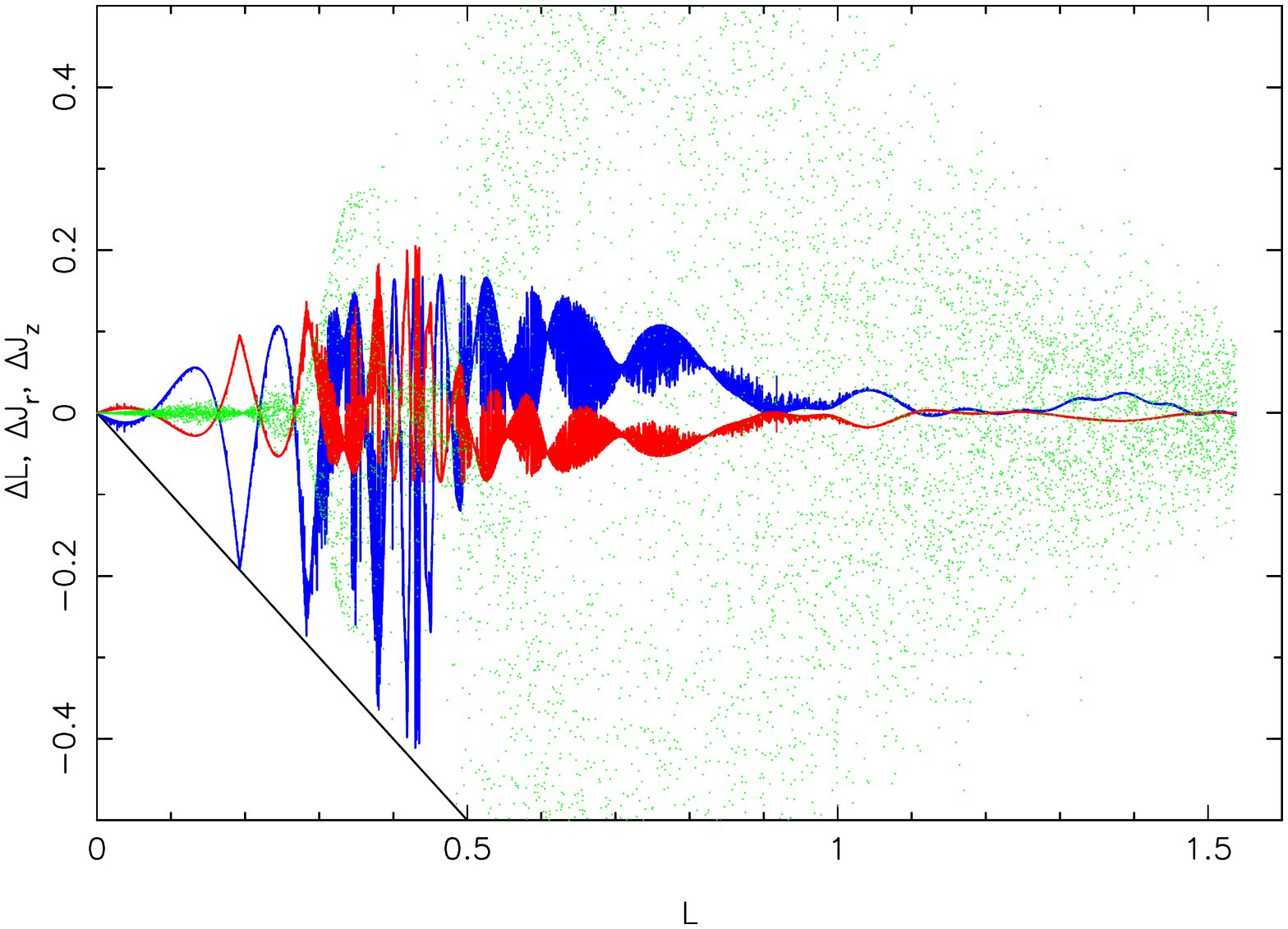} 
\end{center}
\caption{Same as Fig.~\ref{fig_iso145} but for $a_y=-0.10$ and $a_z=-0.05$.}
\label{fig_iso144}
\end{figure}

\begin{figure}
\begin{center}
\includegraphics[angle=-90, scale=0.8]{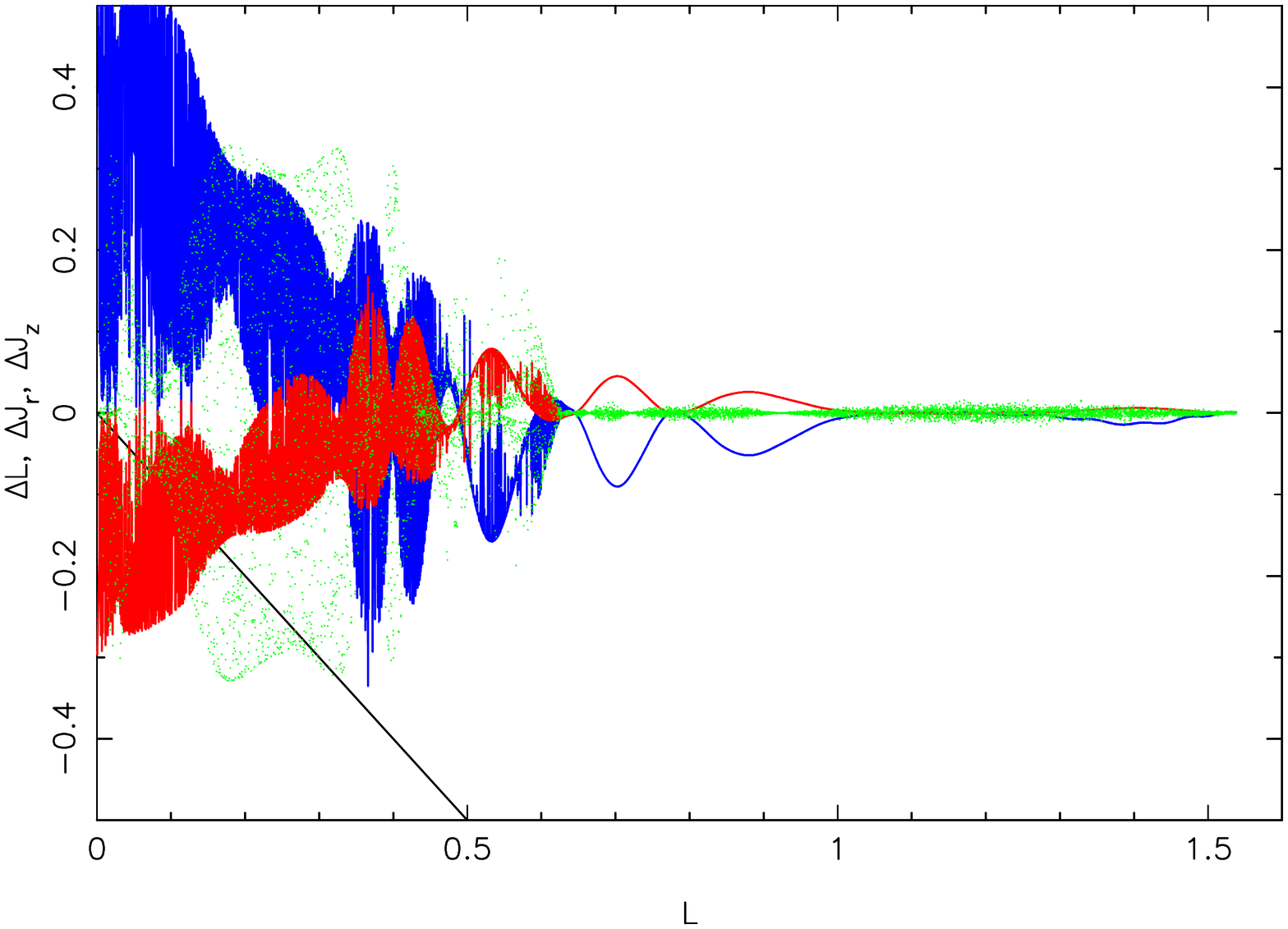} 
\end{center}
\caption{Same as Fig.~\ref{fig_iso145} but for $a_y=0.05$ and $a_z=0.10$.}
\label{fig_iso146}
\end{figure}

\begin{figure}
\begin{center}
\includegraphics[angle=-90, scale=0.8]{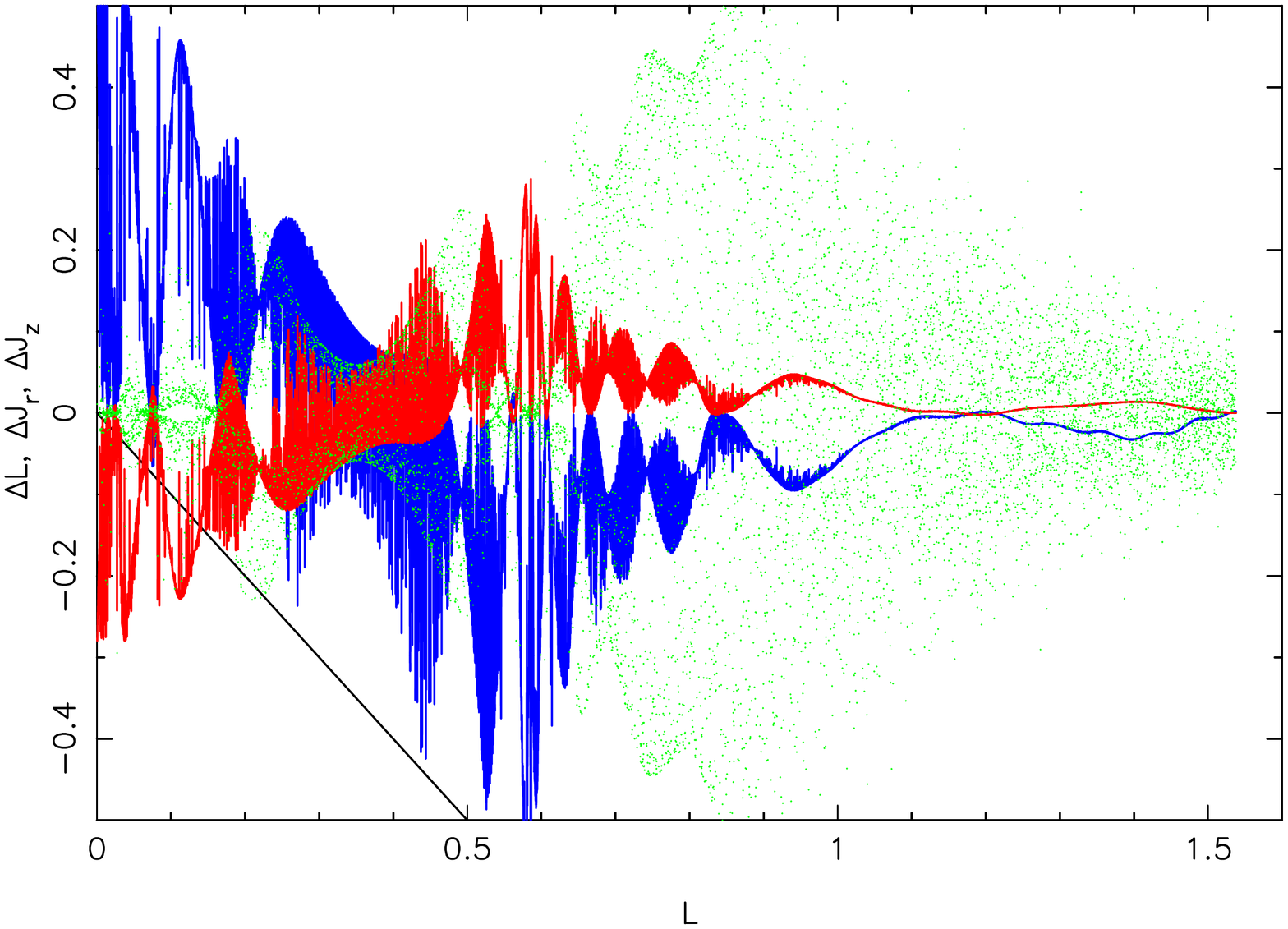} 
\end{center}
\caption{Same as Fig.~\ref{fig_iso145} but for $a_y=0.10$ and $a_z=0.05$.}
\label{fig_iso147}
\end{figure}

\begin{figure}
\begin{center}
\includegraphics[angle=-90, scale=0.8]{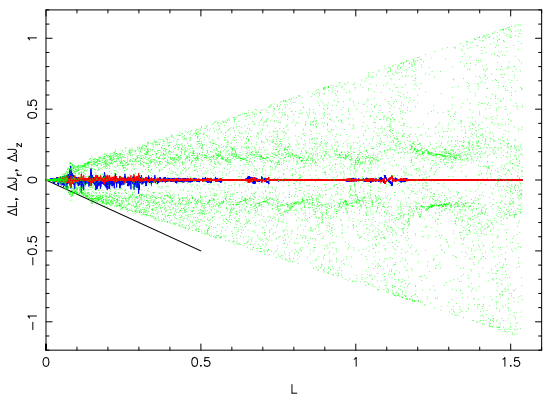} 
\end{center}
\caption{Same as Fig.~\ref{fig_iso144}, with an increased vertical axis,
but for $t_r=2400$, which is 30 times slower than the standard 
rate of potential change. The green dots are the changes in $J_z$. }
\label{fig_iso154}
\end{figure}

\subsection{Test Particle Orbits Without Sub-halos}

Simple orbit integrations in the model potentials illustrate some of the potential's
basic properties. We start a set of 10,000 particles with no 
self-gravity at $x=4$, $y=z=0$ with a uniform distribution in $v_y$ between
zero and the circular value at the chosen starting radius.  The value of $x= 4$ is chosen to help avoid the
core region where there is little effective tidal force.
The particle $v_x$ and $v_z$ velocities are  given a small Gaussian random value, 0.002 units,  about 0.5\%
of the circular velocity,  to allow the particles to explore mildly off
plane orbits. 
The orbits are integrated with a leapfrog integrator
 with step sizes of 0.004 units, which is nearly $10^4$ steps per orbit, conserving
the actions to better than one part in $10^6$. 

Orbits are evolved in 
four different versions of the triaxial isochrone, with $a_y,a_z$ values of $\pm0.05$ and $\pm 0.1$.
The differences of the actions of the particles calculated at the start and end of the orbit integrations are plotted
as a function of the initial orbital angular momentum in 
Figures~\ref{fig_iso145}, \ref{fig_iso144}, \ref{fig_iso146} and \ref{fig_iso147}.
Figure~\ref{fig_iso145} shows changes roughly as one might expect as a function 
of the orbital angular momentum (green points).
That is, the range of initial orbital angular momentum  means that
particles have different orbital phasing relative to the
changing axes. 
Some particles experience a net positive torque and others where the torques lead
to a net loss. 
At high $L$, hence low eccentricity,   the torques largely 
cancel out, leaving little net change. 

In contrast to the simple behavior of Figure~\ref{fig_iso145}, Figures~\ref{fig_iso144}, 
\ref{fig_iso146} and \ref{fig_iso147} show nearly chaotic behavior over a range of orbits \citep{SGV:08}.
The change in potential axis length remains the same
as in Figure~\ref{fig_iso145}, but with the $yz$ axes switched, Fig.~\ref{fig_iso144}, or,  made positive,
hence allowed to expand, Figs.~\ref{fig_iso146} and \ref{fig_iso147}. 
The figures show that nearby orbits have
dramatic differences in the changes to their action variables for higher eccentricity orbits. Low eccentricity orbits at
high angular momentum are generally much less affected, although the vertical action can still be dramatically affected.
This behavior is often seen for orbits that have higher order resonances between orbits and the rate of potential shape change. 

If the potential changes are made very slowly we expect that the action changes for regular orbits will become adiabatic and 
become exponentially small \citep{BT:08}. 
In Figure~\ref{fig_iso154}, the $\Delta L$ and $\Delta J_r$ actions diminish as expected but the changes 
in the $\Delta J_z$ values remain large. 
Overall, the rich range of orbital behavior  of the triaxial isochrone 
makes it a simplified, but interesting, stand-in for a cosmological galactic halo potential that can capture 
realistic star stream behavior beyond a spherical halo.

\subsection{Test Particle Orbits With Sub-halos}

\begin{figure}
\begin{center}
\includegraphics[angle=-90, scale=0.8]{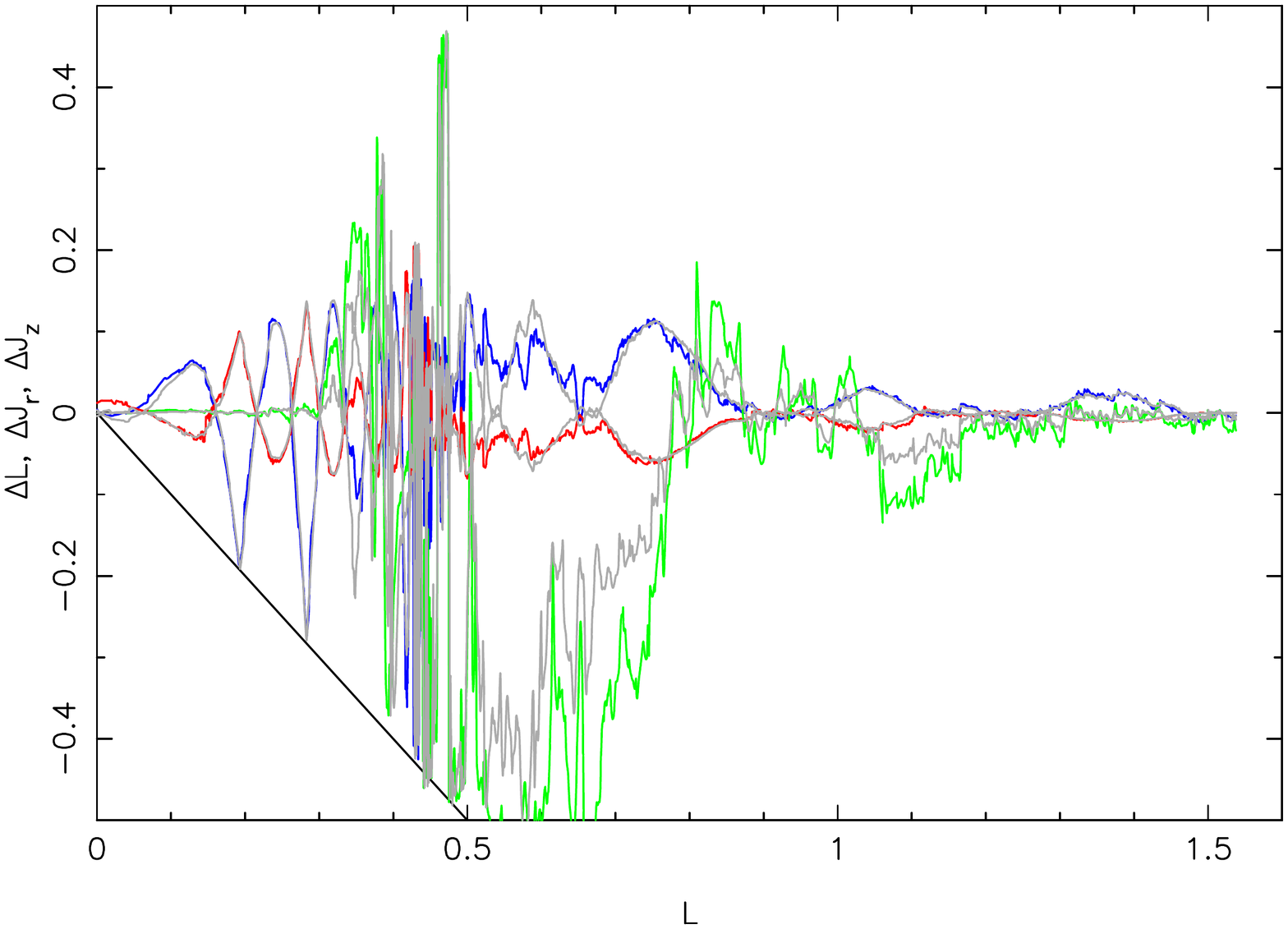} 
\put(-100,-25){\includegraphics[angle=-90, scale=0.25]{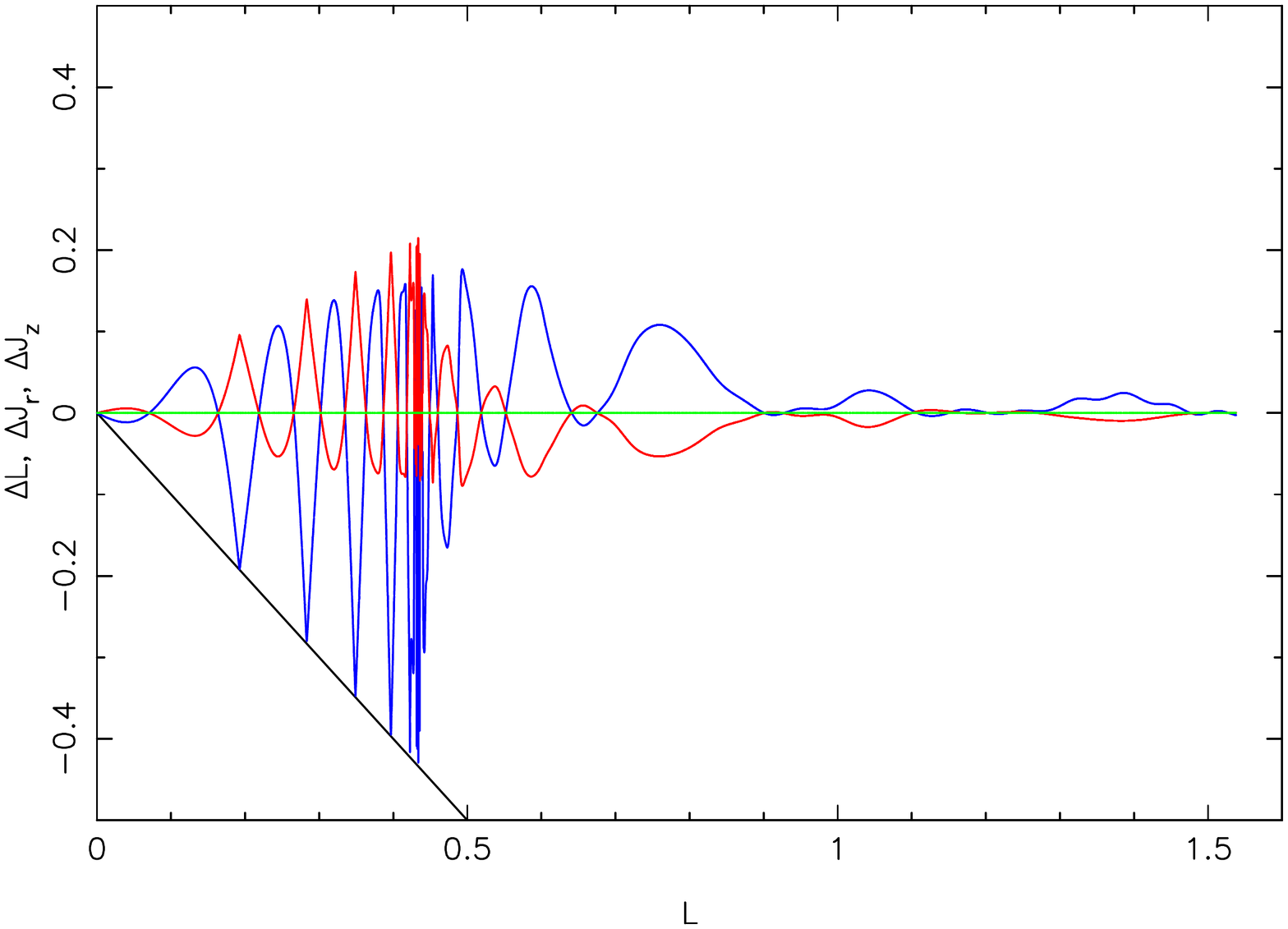}}
\end{center}
\caption{Orbit integrations for the potential of  Fig.~\ref{fig_iso144} with
1000 sub-halos that total to 1\% of the halo mass. The grey
lines show the same simulation with the sub-halos reduced in mass by two. 
 Here the particles are initiated with no additional random velocities.
The inset shows the same simulation with no sub-halos.
 }
\label{fig_iso158}
\end{figure}

The effects of adding sub-halos to the simple orbit integrations of 
Figure~\ref{fig_iso144}, are shown in Figure~\ref{fig_iso158}, where
1\% of the halo mass transfers to 1000 identical mass sub-halos. That is, each
sub-halo has a mass $10^{-5}$ of the host halo. The test particles
are initiated with no additional random velocities to avoid the confusion of orbit diffusion. 
The outcome with no sub-halos is shown as the inset in Figure~\ref{fig_iso158}.

The sub-halo interactions occur on the scale of the radius of the sub-halos, which is approximately 0.02 units 
for the case studied. 
The sub-halos act to create
fairly abrupt changes in the action variables,  spread out over the orbits in accord with the 
density profile of the halo. 
Reducing the mass of each sub-halo by a factor of two, which reduces their scale radius by $\sqrt[3]{2}$,
we find that the sharp changes in $J_z$  for  $L\gtrsim 0.8 $ are one half the height of
the higher mass sub-halo results, indicating that single sub-halo encounters dominate for 
test particle orbits in this range over the duration of the simulation. For $L\lesssim 0.8$ the particles encounter sub-halos
multiple times leading to much larger $J_z$ changes. The highest eccentricity orbits, those with $L\lesssim 0.3$ are largely
unperturbed by sub-halos, likely because of their high velocities through the central regions.
Overall the 
amplitude of the changes in the actions due to sub-halos are comparable to the changes due to 
the triaxial potential.  We can 
foresee that disentangling sub-halo effects from potential effects will not be straightforward as might be hoped.

\section{Simulation of a Small Star Cluster}

\begin{figure}
\begin{center}
\includegraphics[angle=-90, scale=0.8]{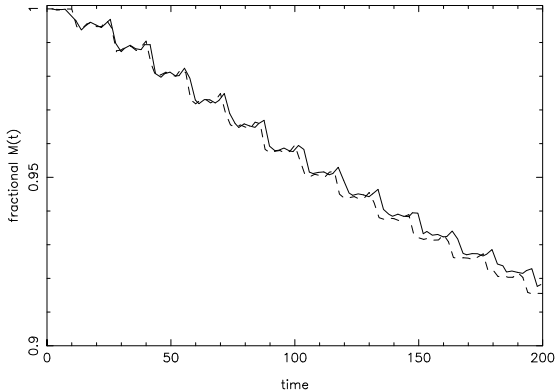} 
\end{center}
\caption{The mass in the cluster as a function of time on an $L/L_c=0.4$ orbit evolved with an $m=10^{-7}$ with a 
full n-body code (solid line) and
with a shell code (dashed line). }
\label{fig_mt}
\end{figure}

\begin{figure}
\begin{center}
\includegraphics[angle=-90, scale=0.7]{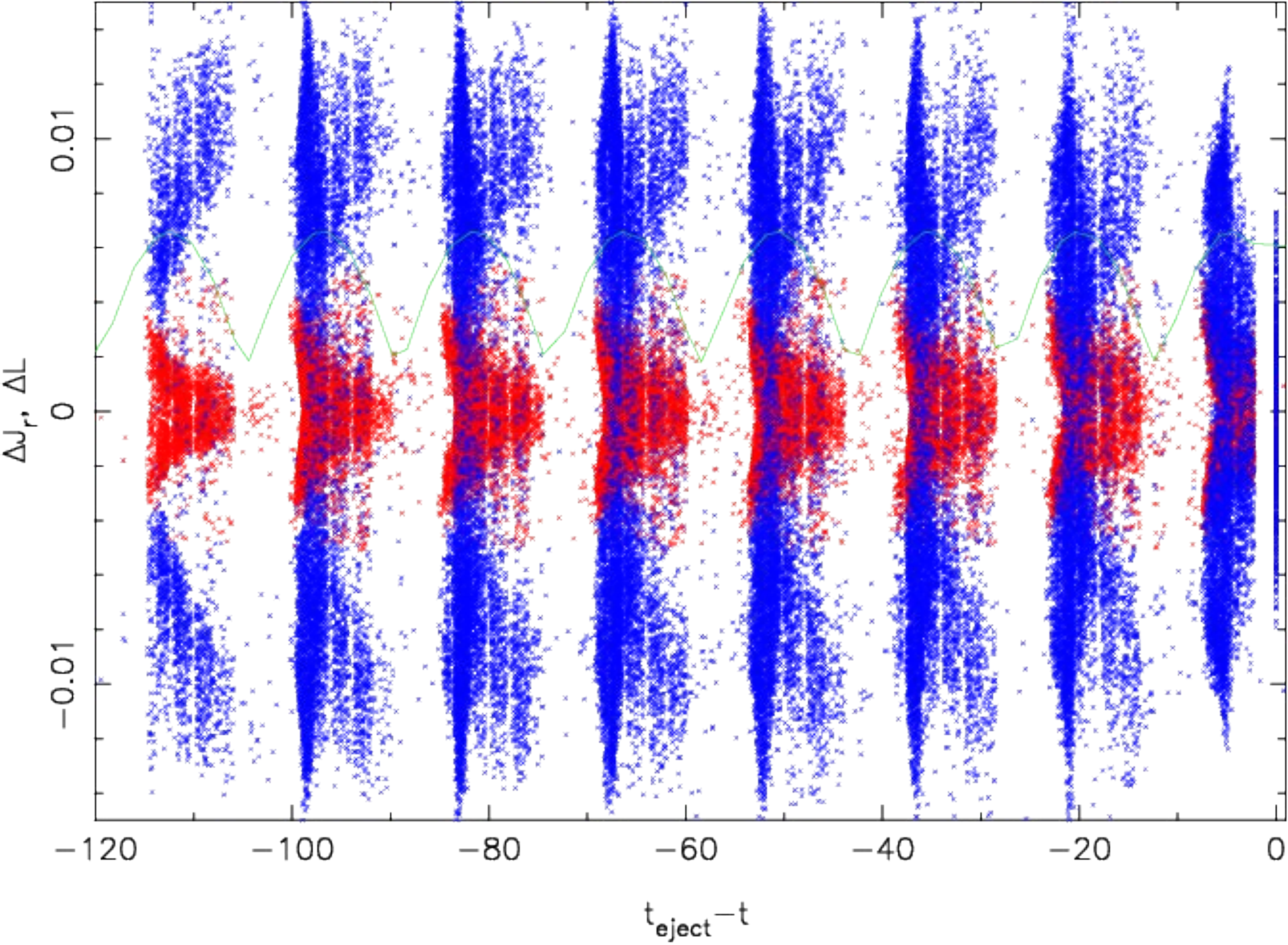} 
\end{center}
\caption{The angular momentum (blue)  and radial action (red) relative to the cluster center, $\Delta L, \Delta J_r$, respectively of particles as a function of their ejection time for the cluster evolved in the Gadget2 n-body code. }
\label{fig_jrtn}
\end{figure}

\begin{figure}
\begin{center}
\includegraphics[angle=-90, scale=0.7]{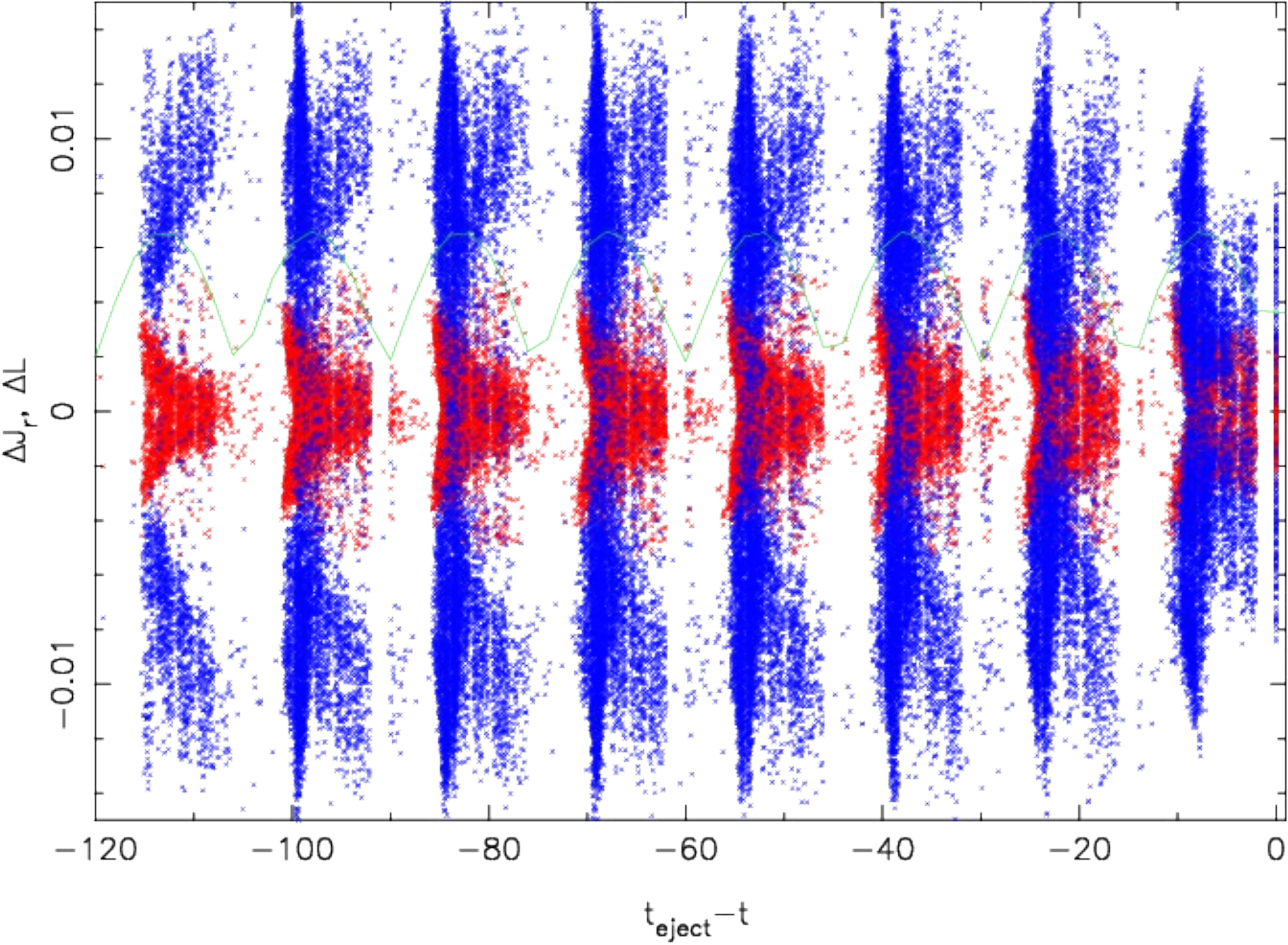} 
\end{center}
\caption{The same model as Fig.~\ref{fig_jrtn} but evolved with the shell code. }
\label{fig_jrts}
\end{figure}

We create a small, self-gravitating satellite of mass of $m \ll 1$ ($m\sim 10^{-8}$, where the host halo has $M_I=1$) 
from a King model distribution function \citep{King:66}. The cluster is created 
with its own dimensionless units of velocity and position, then scaled down in size to fit its zero density 
radius within the 
estimated tidal radius. The velocities  are scaled to the desired mass and new size.
Our standard star cluster model
has a central potential, $\psi(0)$, relative to the  central velocity dispersion, $\sigma$, of 
$-\psi(0)/\sigma^2=4$. Increasing $-\psi(0)/\sigma^2$  decreases the
core radius relative to the  radius where the model density goes to zero. 
The outer density profile has a slope nearly independent of the core radius, 
so the mass loss into the tidal stream has a weak dependence on the 
concentration of the model star cluster.
Consequently the star cluster's core radius only begins to have a significant effect on the
tidal stream when mass loss is so large
that the core is eroded.

The outer radius of the star cluster model
is  scaled to be equal to the tidal radius, $r[m/(3M_I(r))]^{1/3}$, where
$r$ is the starting radius of the orbit. 
The small
adjustment to the tidal radius for a distributed mass potential \citep{BT:08} is not included, nor is any allowance
for a non-circular orbit, 
which leads to some modest internal changes of the star cluster model
 in response to the tidal field in its first orbit. 

The standard model uses $10^5$ particles in the initial star cluster.
The star particles in the model star clusters are subject to gravitational forces from other star particles, the 
host galaxy gravitational potential, and any orbiting sub-halos that may be present.
The rate of mass ejection varies around the orbit and 
leads to complex, strongly correlated, velocity differences (or action variable offsets) between the stream and the star cluster
\citep{Carlberg:15}.   The mass loss in the
systems of interest is driven by tidal heating and is well modeled  with a collisionless n-body code. 
As a reference, we use the 
gravitational and orbital evolution components of the well known and characterized Gadget2 code \citep{Springel:05}.

\subsection{An n-body shell code}

The precise description of the evolution of a star cluster in the tidal field of a host galaxy remains an important
research problem as a generalization of the dynamics of an isolated
star cluster.  Current work allows for a realistic range of star masses, their mass evolution and their remnants, 
the presence of binaries and higher multiplicity systems.
The evolution of collisional star clusters in tidal fields generally either use some form of gravitational
calculation splitting to handle the
dynamics on different time scales \citep{Fujii:07} or interactions via statistical approximations \citep{SMB:14,Brockamp:14}.
\citet{FH:00} demonstrate the 
complexity of star cluster orbits near the tidal radius that exists for cluster in a circular orbit.

Two well known thin star streams are Pal~5, which emanates from an exceptionally low concentration cluster 
\citep{Odenkirchen:01} and GD-1, which has no 
known progenitor \citep{GD:06}. Given the complexity and high computational costs
of studying a collisional star cluster, and, the observational fact that the importance of star-star
encounters in the progenitor clusters is either relatively low, or, unknown, 
it is currently conventional to study tidal streams as emanating from tidally heated collisionless star clusters \citep{Dehnen:04}.
Given that our interest is in the subsequent dynamics of the star streams the assumption is appropriate for
the time being, but it will eventually be important to incorporate a more complete dynamical description of the
originating star cluster.

The evolution of a very low mass, collisionless,  satellite orbiting in an external potential in which dynamical friction is not a
factor is well handled with  multipole expansion codes. The relatively low cost and good accuracy 
of multipole n-body codes  has a rich history
as the first codes that could handle large numbers of
self-gravitating collisionless particles 
\citep{Aarseth:67,vAvG:77,FP:80,Villumsen:82,White:83,McGlynn:84}.
Multipole methods remain
popular for moderately distorted spheres with arbitrary density profiles
for which there are now efficient parallel processing algorithms \citep{Meiron:14} and
have been generalized to the powerful self-consistent field method \citep{HO:92}.
A multipole field provides substantial smoothing of 
the gravitational field and completely suppresses the complex astrophysics of the two and more body 
interactions in a dense globular cluster.  
If desired, star-star collisional physics can be added using a Monte Carlo method \citep{Giersz:06}
although none is included here.

After testing potentials calculated with up to fourth order multipole
expansions we found that a simple shell code, which only calculates the 
spherical part of the gravitational field,
 gave results that were essentially indistinguishable, with the shell code being
less subject to the problem of defining the center. 
This is not unexpected in the situation
we are modeling since much of the mass of the satellite is in the central regions where the tidal field
is weak. On the other hand, the degree of quantitative accuracy needs to be verified.
The shell code calculates the radial gravitational acceleration of a particle at radius, $r$ 
with a mildly softened gravity, $-GM(<r)r/(r^2+s^2)^{3/2}$ where $M(<r)$ is simply the total mass interior to  $r$  relative to a central point.  
The softening $s$ is set at 20\% of the core radius of the King model cluster we wish to simulate and is needed for the case 
where two particles are near the coordinate center.
Making a good choice for the central point of the potential is important, since a poor choice can lead to
undesirable oscillations and even 
an unstable cluster. For the situations of interest here the simple approach of integrating 
the orbit of a fictitious particle placed at the position of the coordinate center of the cluster works extremely 
well and shows no signs of instability. 

The calculation sorts particles in order of ascending radii and then calculates the gravitational field
at each successive particle, simply as the mass interior to its radius. The sub-halo gravity is calculated
with a direct sum.
Our shell code has fixed time steps, $\Delta t=0.004$,
 that accurately capture the tidal heating in the star cluster and the interactions with sub-halos down the tidal stream.
For many of our simulations about half of the particles remain in the cluster so increasing the time step size for particles 
outside the cluster would not necessarily give rise to a large speed increase. 
The code was parallelized with a parallel quicksort and MPI routines to reduce the execution time. Other than the quicksort
the algorithm is essentially trivially parallel.

\subsection{Comparison of n-body and shell code results}

As tests of the accuracy of the shell code,  Figure~\ref{fig_mt} shows the mass as a function of time in
Gadget2 and the shell code
for a cluster started at $r=2.2$ in a spherical isochrone. 
Note that difference between the full n-body code and the shell 
code is almost entirely one of slightly different orbital periods, not of mass lost per orbit. The shell code uses
the high accuracy orbit of the center of the star cluster.
The angular momentum and the radial action calculated at the end of the simulation when 
the potential is again spherical are measured relative to the star cluster as a function of their time of ejection from 
the cluster.  The resulting distributions  are shown for the n-body simulation in
Figure~\ref{fig_jrtn} and for the shell code in Figure~\ref{fig_jrts}. 
The two are essentially identical, other than the slight difference in the orbital period, with the shell code 
having the correct period.  
We conclude that the shell code  gives accurate results for the range of  satellites, orbits and potentials of interest here. 

\begin{figure}
\begin{center}
\includegraphics[angle=-90, scale=0.65]{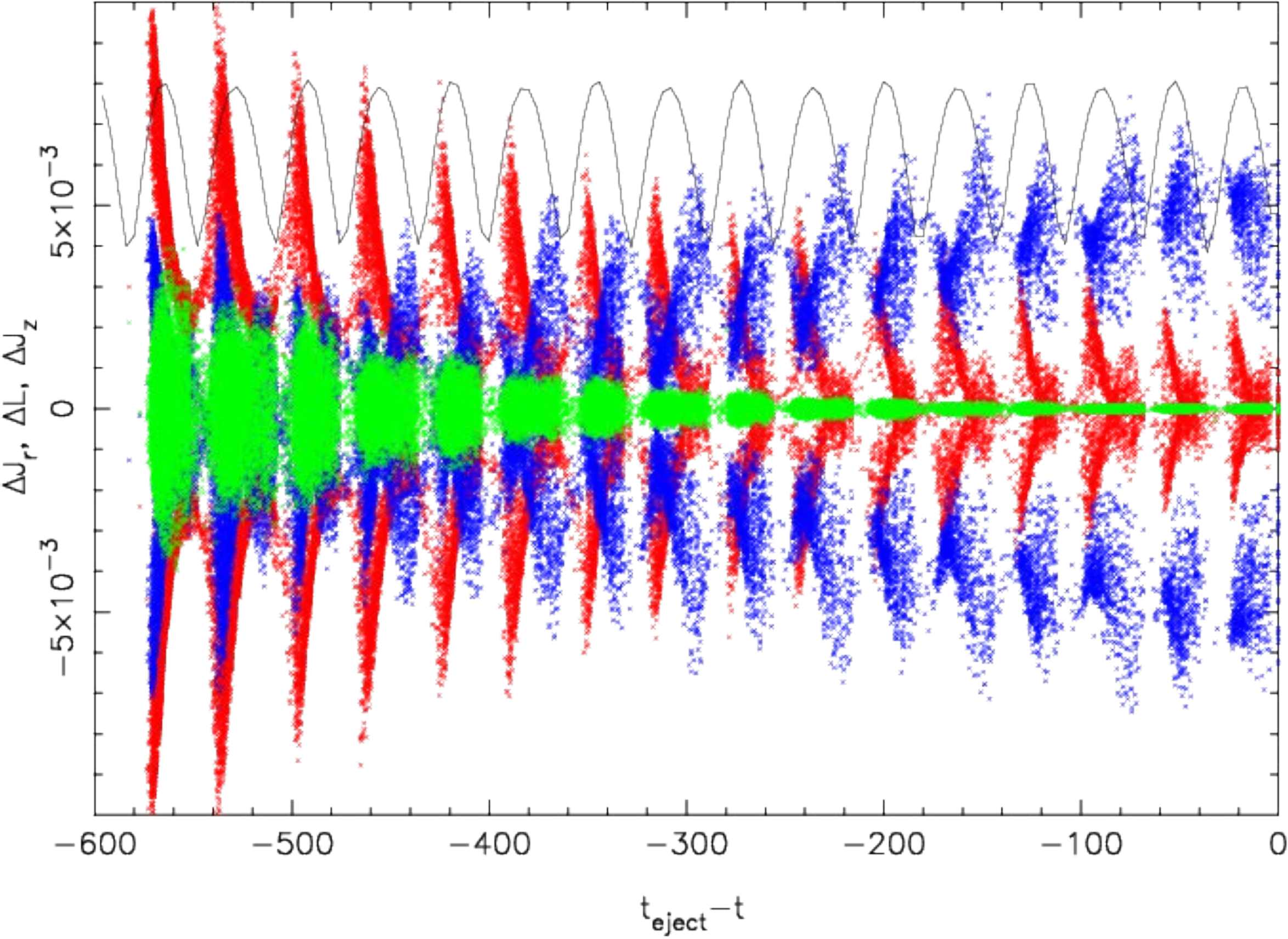} 
\includegraphics[angle=-90, scale=0.65]{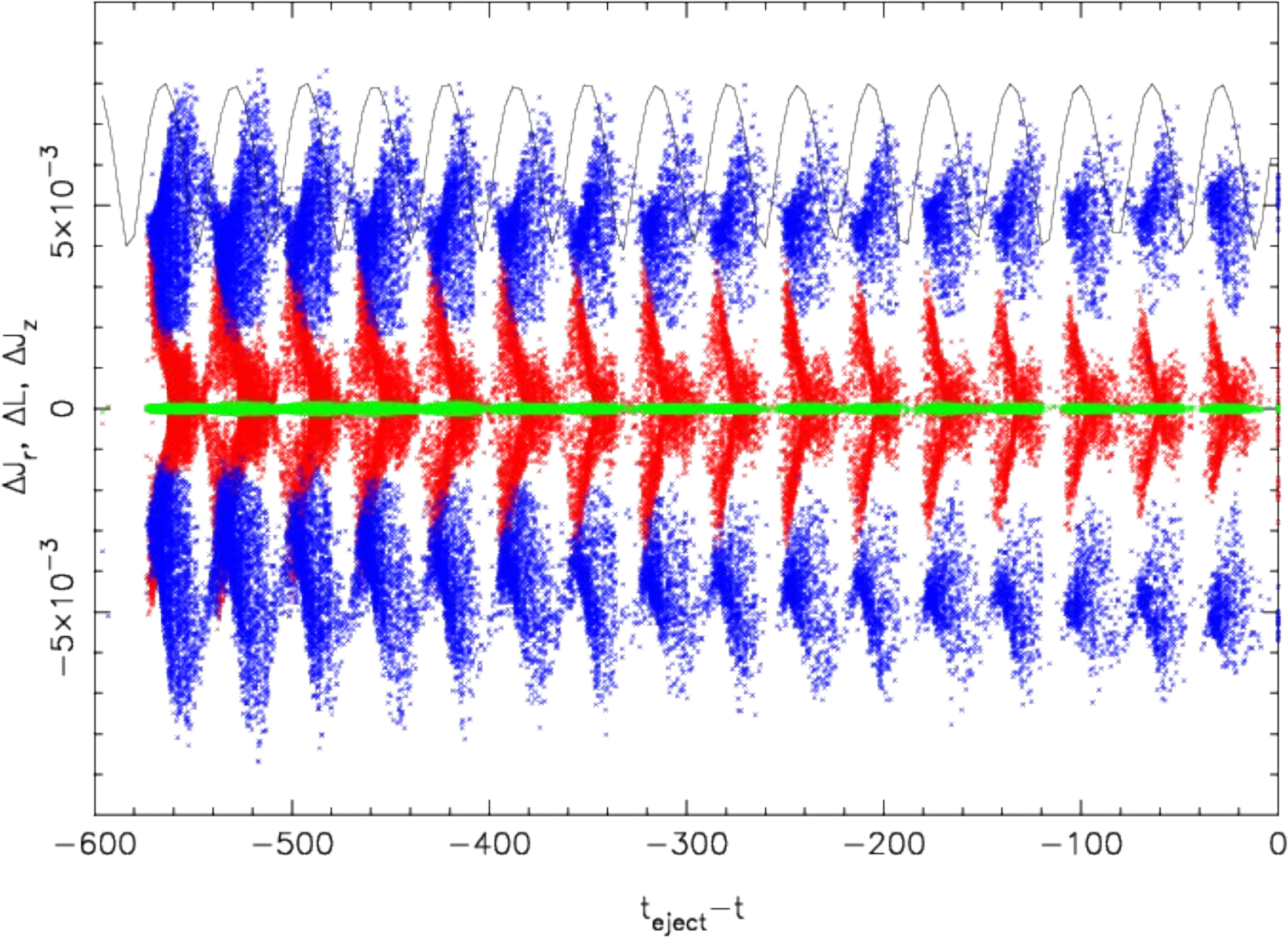} 
\end{center}
\caption{The changes of the actions of particles of a stream from a cluster started with $L/L_c=0.7$ in a smooth triaxial with
$[a_y,a_z]=[-0.1,-0.05]$ (top) and $[-0.05,-0.1]$ (bottom) The actions are measured at the end of the simulation in 
the spherical potential  relative to the cluster center values.
The differences are plotted as a function of the time since the particle was
ejected from the cluster. 
The action variables are red for $\Delta J_r$,   blue for angular momentum, $\Delta L$ and green
for $\Delta J_z$ whose values are scaled downward by a factor of 30 in this plot. }
\label{fig_jrpz7}
\end{figure}

\begin{figure}
\begin{center}
\includegraphics[angle=-90, scale=0.65]{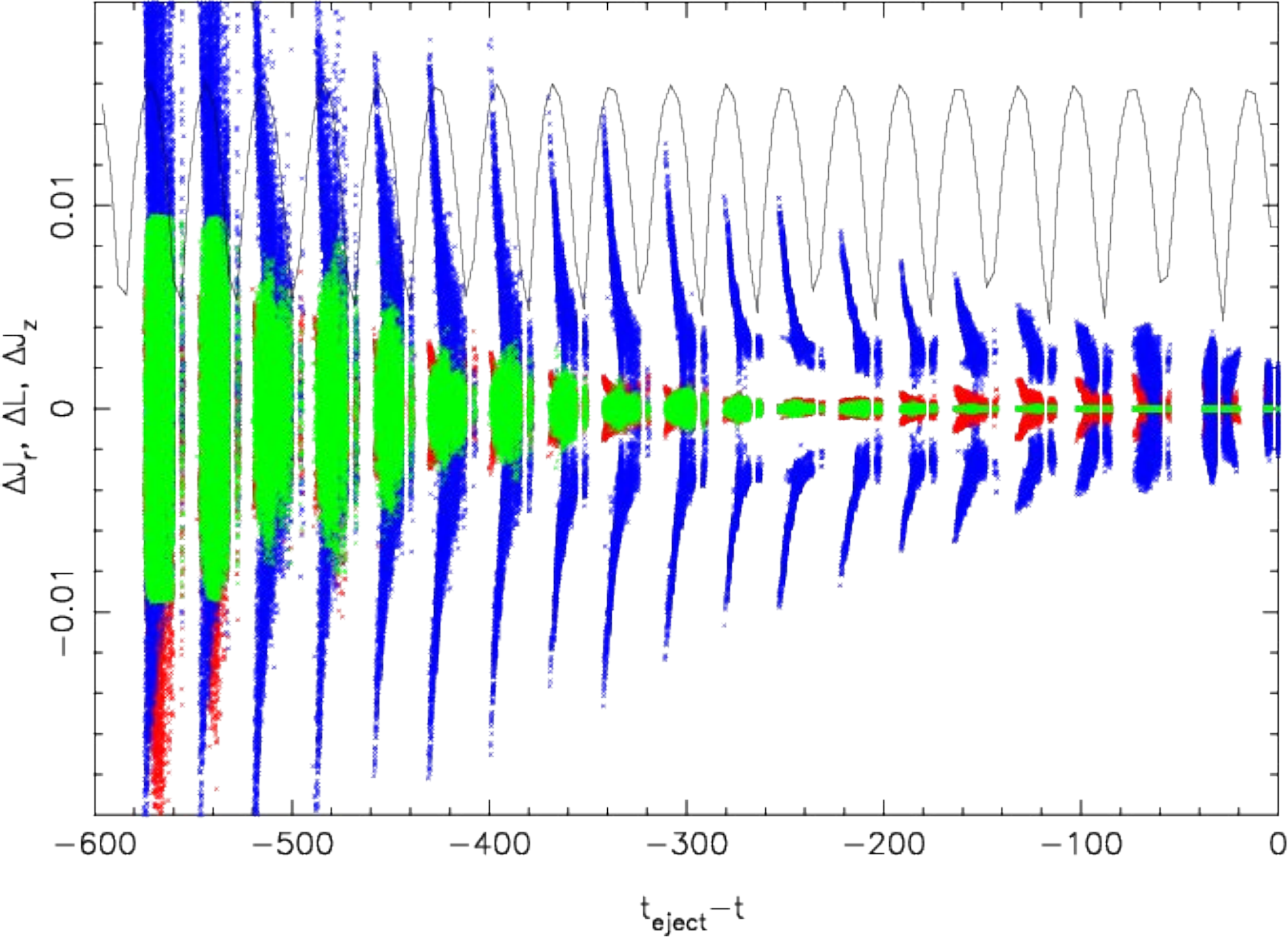} 
\includegraphics[angle=-90, scale=0.65]{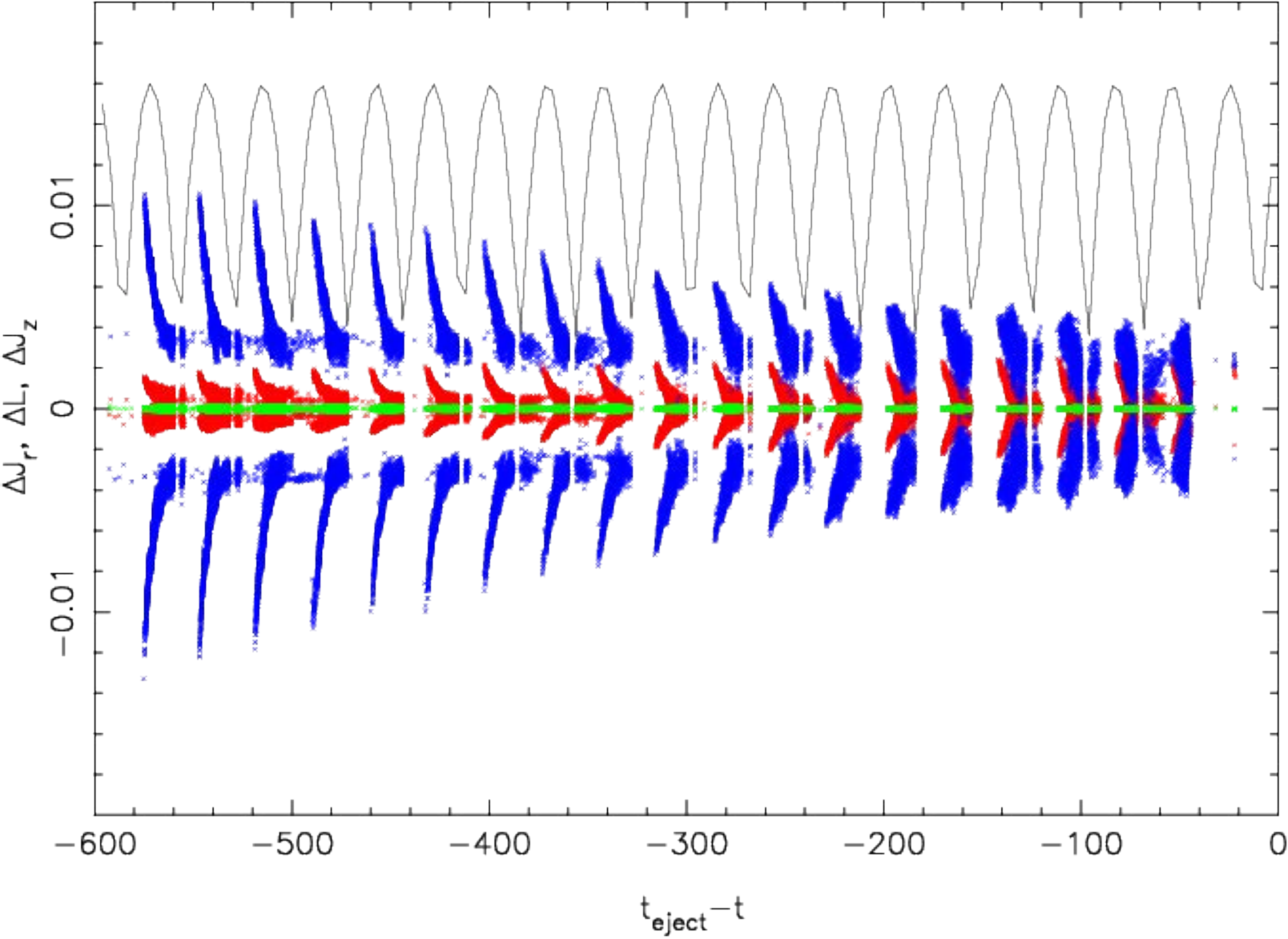} 
\end{center}
\caption{Same as Figure~\ref{fig_jrpz7} but with a cluster started with $L/L_c=0.4$. Note that the vertical scale is reduced
relative to Fig.~\ref{fig_jrpz7}. }
\label{fig_jrpz4}
\end{figure}

\subsection{Streams  in Triaxial Potentials without Sub-Halos}

Clusters were started at $x=4$ with
$L/L_c=0.7$  and $0.4$  in  spherical and  $[a_y,a_z]=[-0.1,-0.05]$  triaxial isochrones with no sub-halos.
A cluster in a low eccentricity orbit 
initiated with $L/L_c=0.7$ loses 39\% of their mass over the interval to time 600 nearly independent 
of triaxiality. The cluster undertakes
approximately 17 radial orbits in the spherical potential.
Over the same time a cluster on a high eccentricity orbit
loses 99.7\% of its mass in the the triaxial potential and 94\% in the spherical potential (21 radial orbits). Although the
difference is easily measurable, when considered as a rate the differences with potential shape
 are small compared to the dependence on 
orbital eccentricity.
The mass loss does depend on the location of the surface used to measure contained mass. We use 0.03 units, approximately
three times the initial tidal radius.

Figures~\ref{fig_jrpz7} and \ref{fig_jrpz4} show the action variables relative to the star cluster of the particles in the stream as function
of their ejection age for two triaxial potentials.
Two clear features are that
the two tidal tails remain almost completely symmetric,  largely a consequence
of the small size of the cluster, which has a tidal radius of 0.007  units,
about 0.3\% of the orbital radius.
Consequently, the tidal field is  essentially symmetric across the cluster and
the two tails are nearly the same, although moving away from the cluster in opposite directions. 
The second feature
is that the radial action and angular momentum of the stars relative to the satellite
 vary slowly  and smoothly over the duration which particles in the stream.
The change of the angular momentum in the initial orbital plane, $\Delta J_z$,
 is largely a measure of the subsequent tilt of the orbits.
$\Delta J_z$ rises
exponentially with time,  the doubling time being approximately 4 radial orbits. The exponential growth of the 
orbit tipping is an important new feature
that the triaxial potential brings into play for the tidal stream relative to an orbit started with the same velocities
in a spherical or axisymmetric potential.  As the lower panels of Figures~\ref{fig_jrpz7} and \ref{fig_jrpz4} show
the same triaxility on different axes relative to the orbital plane does not show the exponential growth. 
Hence, orbital structure will depend on the details of the potential and the initial orbits of particles in the potential.

\section{Streams  in Triaxial Potentials with Sub-Halos}

\begin{figure}
\begin{center}
\includegraphics[angle=-90, scale=0.65, clip=true, trim=45 30 30 40]{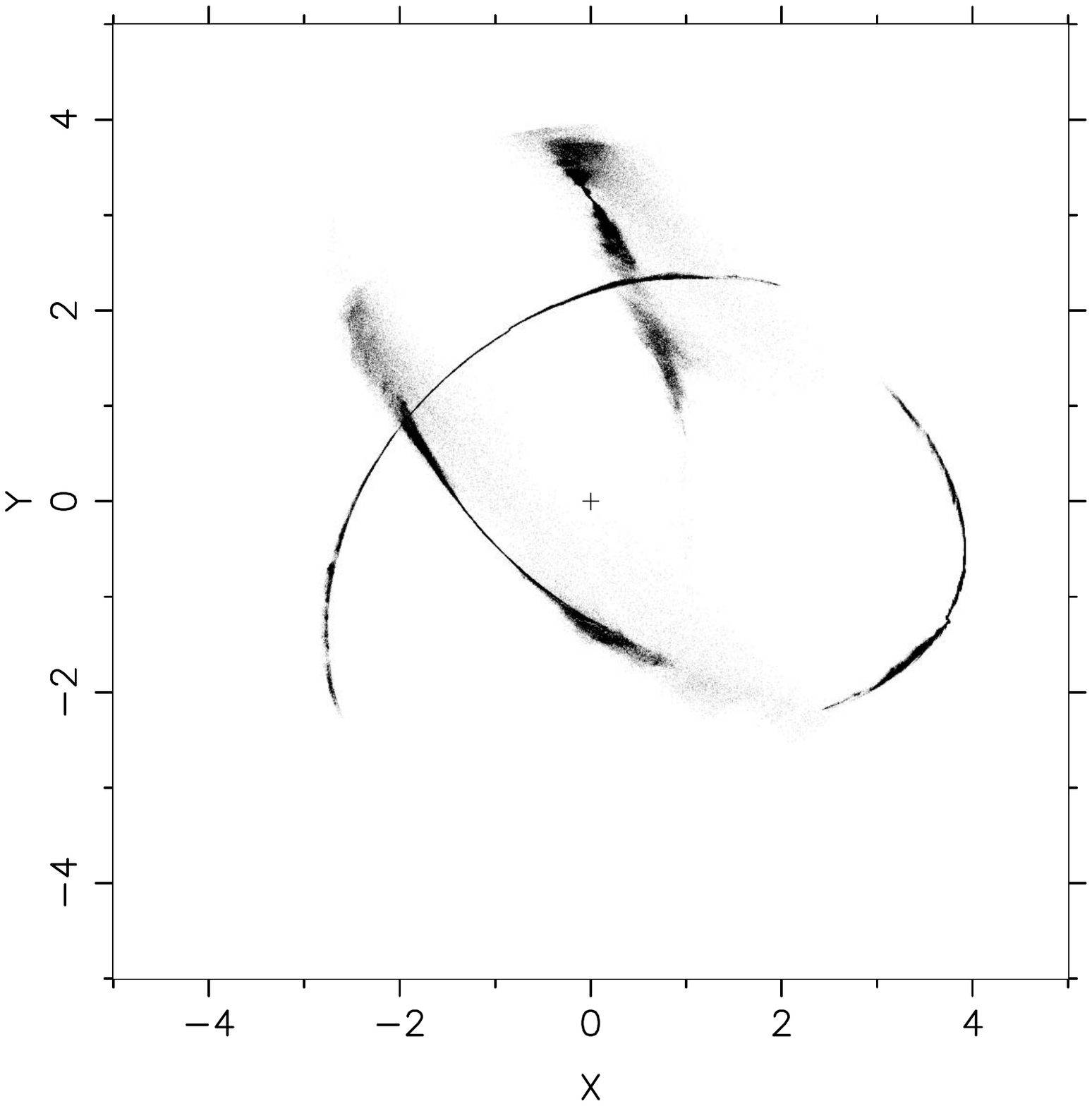} 
\includegraphics[angle=-90, scale=0.65, clip=true, trim=45 30 30 40]{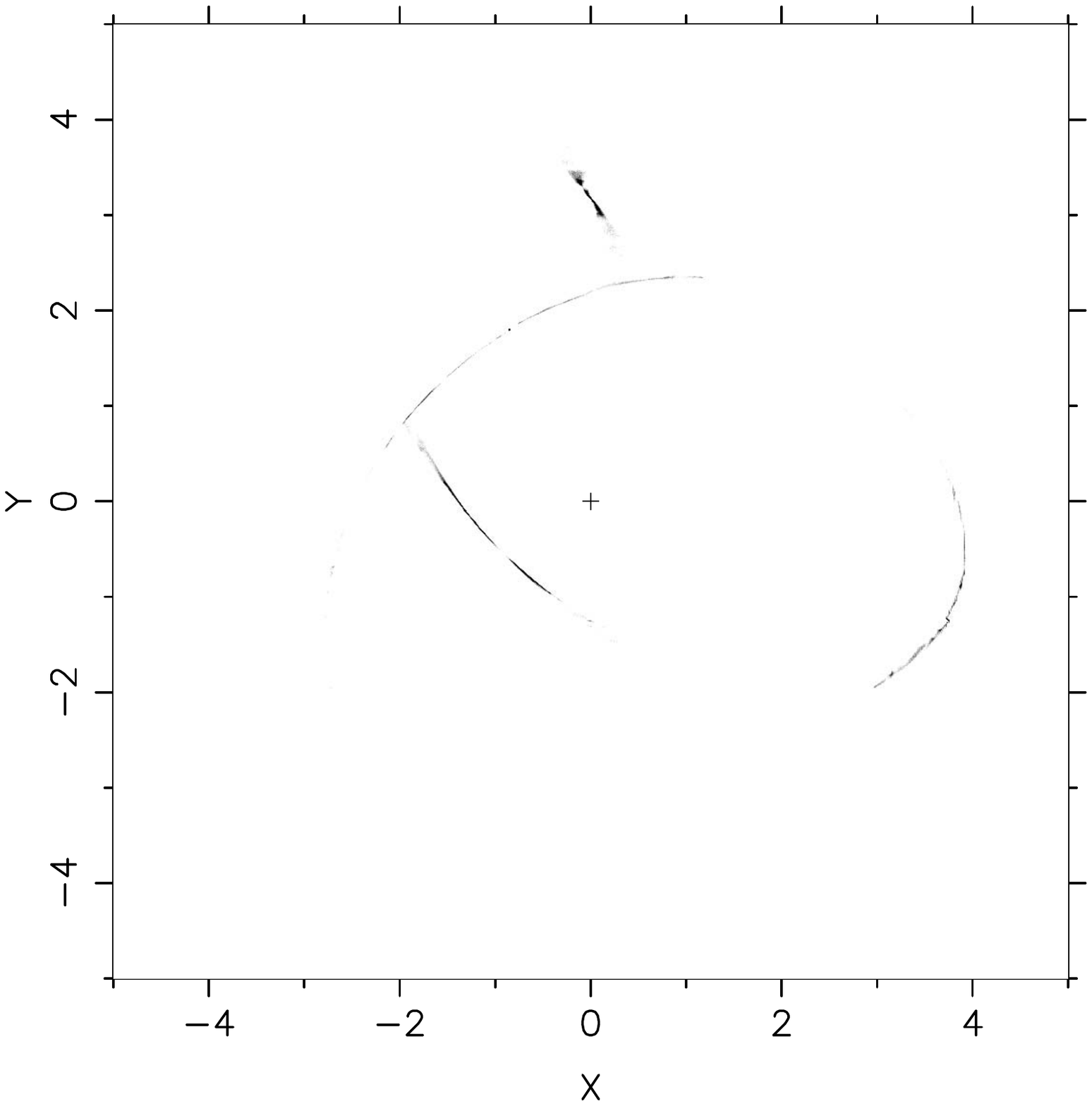} 
\end{center}
\caption{Snapshots at times 580 and 600  of a low and high eccentricity stream. 
In the top panel the low eccentricity stream is the two blurry structures that come near the center, 
whereas the high eccentricity stream is the thinner structures that cross them and nearly connect.
The contrast scale goes from 0 to 10 particles per 0.01 unit square pixel in the top panel
and from 20 to 100 particles per pixel in the bottom.
}
\label{fig_xyorb}
\end{figure}

Two star clusters started with the same velocites in the $a_y=-0.1$, $a_z=-0.05$ halo of Figures~\ref{fig_jrpz4} and
\ref{fig_jrpz7}, but now with 1000 sub-halos 
containing 0.5\% of the 
mass of the halo, is shown in projection (both at two times) onto the orbital plane in Figure~\ref{fig_xyorb}. The addition
of the sub-halos leads to gaps in the stream and helps to push the progenitor and stream
onto a somewhat different orbits relative to the same potential without sub-halos.

\subsection{Stream Visibilities}

\begin{figure}
\begin{center}
\includegraphics[angle=-90, scale=0.8]{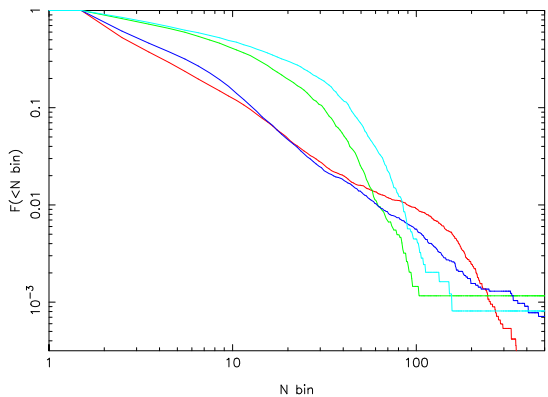} 
\end{center}
\caption{The cumulative distribution, normalized to the total,
 of surface densities in the low eccentricity stream (green and turquoise lines) and the high 
eccentricity stream at times 580 and 600 (red and blue lines) in the $[-0.1,-0.05]$ halo shown in Fig~\ref{fig_xyorb}.}
\label{fig_nl47}
\end{figure}

A star cluster on a high eccentricity orbit is subject to stronger tidal fields at pericenter than a low eccentricity cluster with
the same apocenter. The high eccentricity orbit also takes the stream into regions where there are more sub-halos, therefore
we expect high eccentricity streams to have more orbital spreading than low eccentricity streams. To examine this
idea we compare the streams from clusters on $L/L_c=0.4$ and 0.7 orbits in Figure~\ref{fig_xyorb}.
A cluster started with $L/L_c=0.4$ loses mass at more than twice the rate of a cluster
started with $L/L_c=0.7$. Over the somewhat arbitrary simulation duration here
the high eccentricity mass cluster is reduced to a few percent of its initial mass and it is only a coincidence of timing that the final moment is caught.
Figure~\ref{fig_xyorb} shows the projected view of the two  streams at two snapshot times.
 The upper panel shows
the location of all particles, showing that the high eccentricity stream has many stars
scattered well away from the stream.  
In currently available observational data, once  match filtered to identify streams,
it usually is difficult to identify stream locations where the local surface density is less
than about a factor of 5 in density below the peak values \citep{CG:13}.

Figure~\ref{fig_nl47} shows that the surface density distribution of high and low eccentricity streams 
have a different character, with low eccentricity streams having relatively more paricles
at higher surface densities. The distribution is not strongly orbital phase dependent.
The peak values are similar, with
about  100 particles per pixel (each pixel is 0.01 units square)  for both low and high eccentricity 
stream,  although we do note that the high eccentricity stream has somewhat higher
peak densities. 
At  a factor of 5 down from peak, 20 particles per pixel, about 25\% of the low eccentricity stream
is visible but only 5\% of the high eccentricity stream.  The lower panel of Figure~\ref{fig_xyorb} 
shows the regions with surface densities above 20 particles per pixel.
More quantitatively, from Figure~\ref{fig_nl47} we see that
 95\% of the high eccentricity stream is below the 20\% of peak surface 
brightness level, nearly independent of the orbital phase.
Similar behavior is seen in the more general simulations of \citet{Ngan:15b}, the results
here clarifying the combined roles of sub-halos and triaxiality in allowing much of the stream to disperse
to low densities.

The visible segment of the Ophiuchus stream 
has a luminosity of $1.4\pm0.6 \times 10^3 \lsun$ \citep{Bernard:14} and 
is estimated to be on an $e=0.67$ orbit \citep{Sesar:15}. 
If the visible section of Ophiuchus
has a similar threshold for visibility as the high eccentricity stream here, then the total luminosity of the entire Ophiuchus 
stream could be 20 times higher, a total of $3\times 10^4 \lsun$
These exploratory ideas mainly point to a scenario to understand the very short and
low luminosity  Ophiuchus stream which needs
to be developed into a model to test its viability.

\section{Stream Action Variables}

Action variables are constants of the motion in a stationary potential.  If the actions have the
same distribution of values along the length of a stream, then they can be used along with
the stream shape and any available velocities to put strong 
constraints on the mass and shape of the galactic dark matter potential. The actions of stars ejected 
into a stream  have pulses which vary systematically around an orbit,  
but these details average together in position space a few orbital cycles further down
the stream \citep{Carlberg:15}.
As seen in Figures~\ref{fig_jrpz7} and \ref{fig_jrpz4} a time variable triaxial potential 
induces smooth systematic changes in the potential which are symmetric relative to the
satellite. Although such a smooth variation will complicate modeling the constraining 
power of streams remains. 

\begin{figure}
\begin{center}
\includegraphics[angle=-90, scale=0.65]{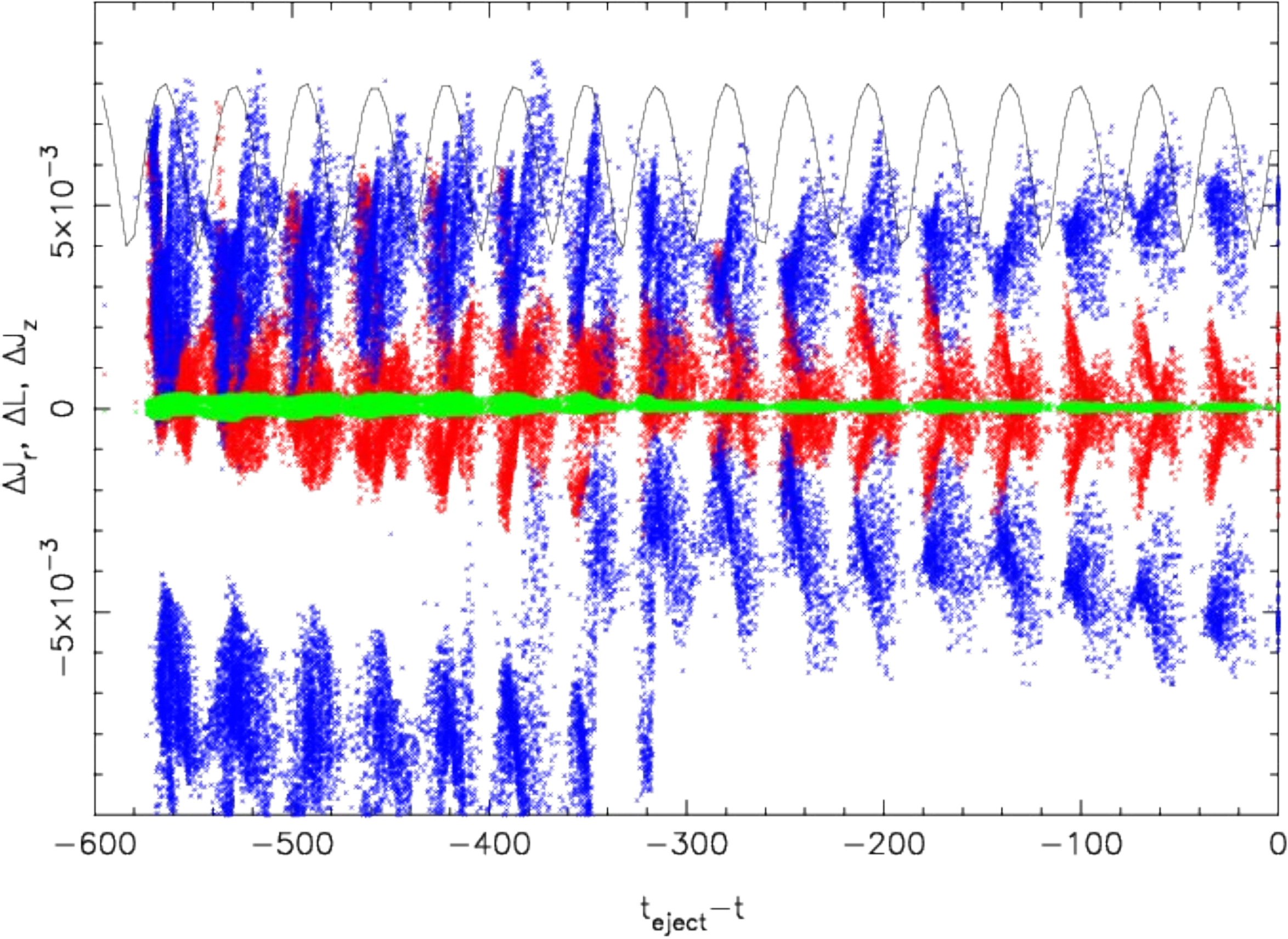} 
\includegraphics[angle=-90, scale=0.65]{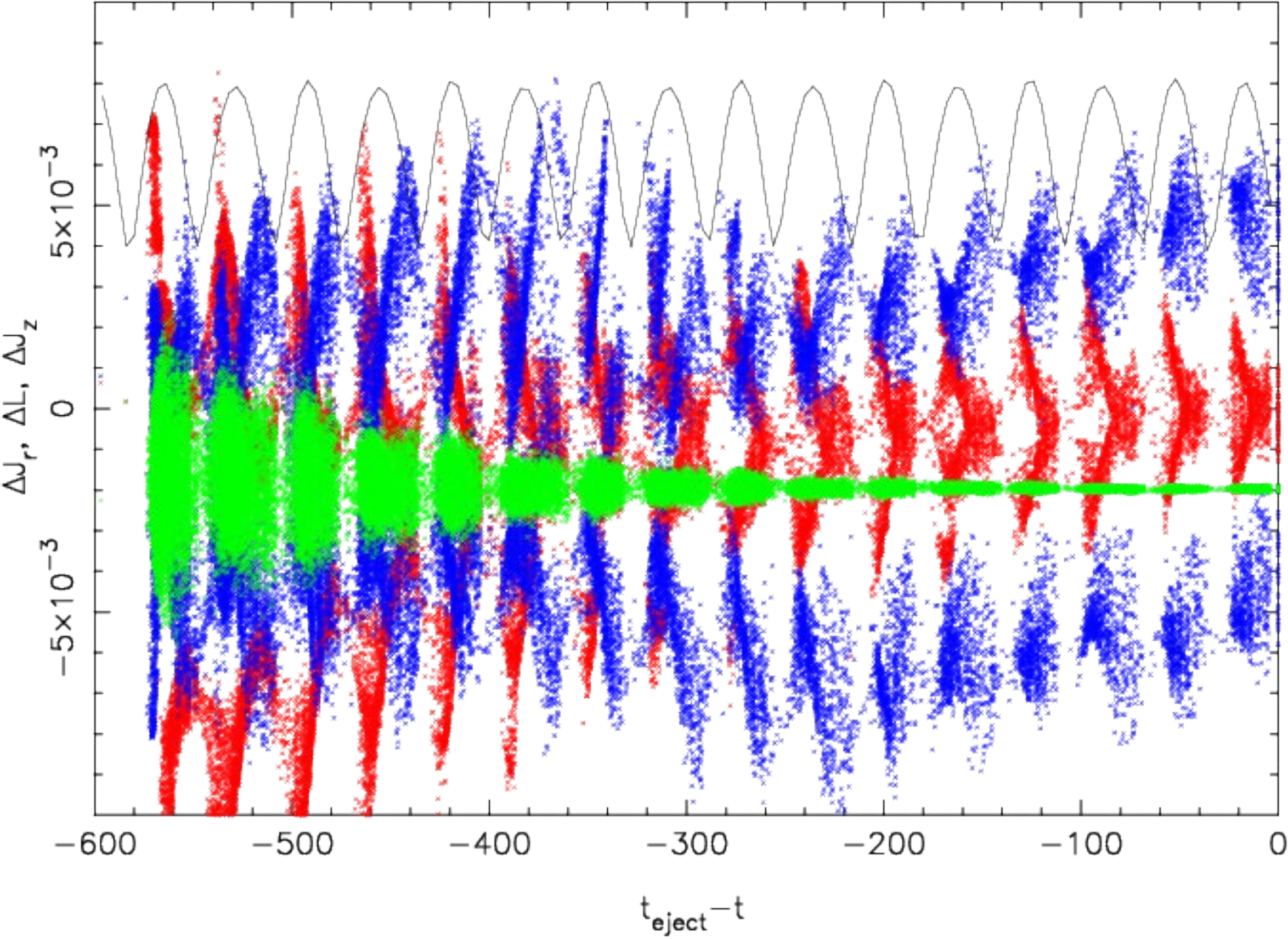} 
\end{center}
\caption{The actions of stream particles relative to the satellite center for the $L/L_c=0.7$ satellite shown
in Fig.~\ref{fig_xyorb}, as a function of ejection time. The triaxial potentials
$[a_y,a_z]=[-0.05,-0.1]$  (top) and $[-0.1,-0.05]$ (bottom)
contain 1000 sub-halos with 0.5\% of the mass.
The $\Delta J_z$ values are scaled down a factor of 30 to fit on the plot. }
\label{fig_jrpz7h}
\end{figure}

\begin{figure}
\begin{center}
\includegraphics[angle=-90, scale=0.65]{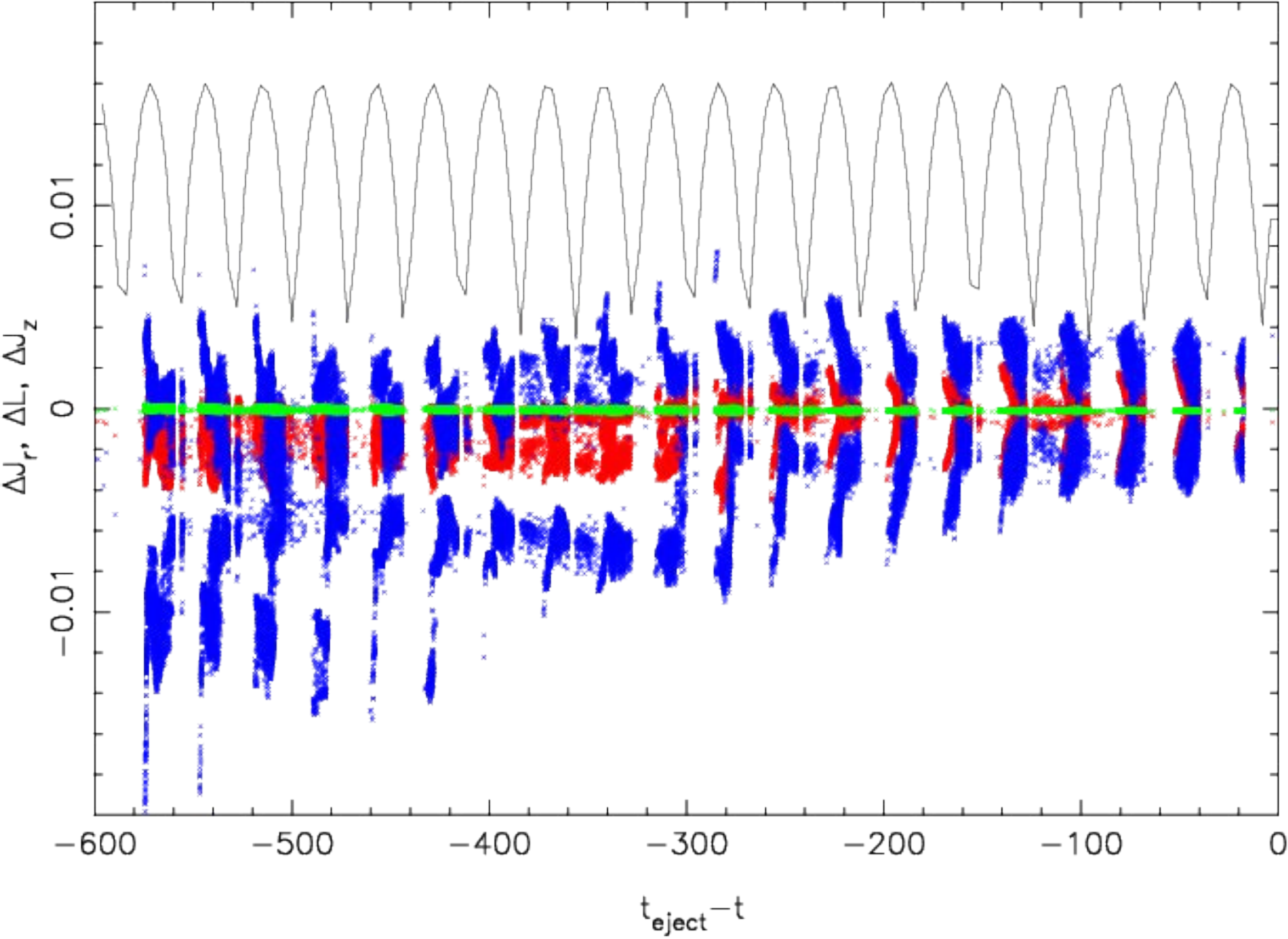} 
\includegraphics[angle=-90, scale=0.65]{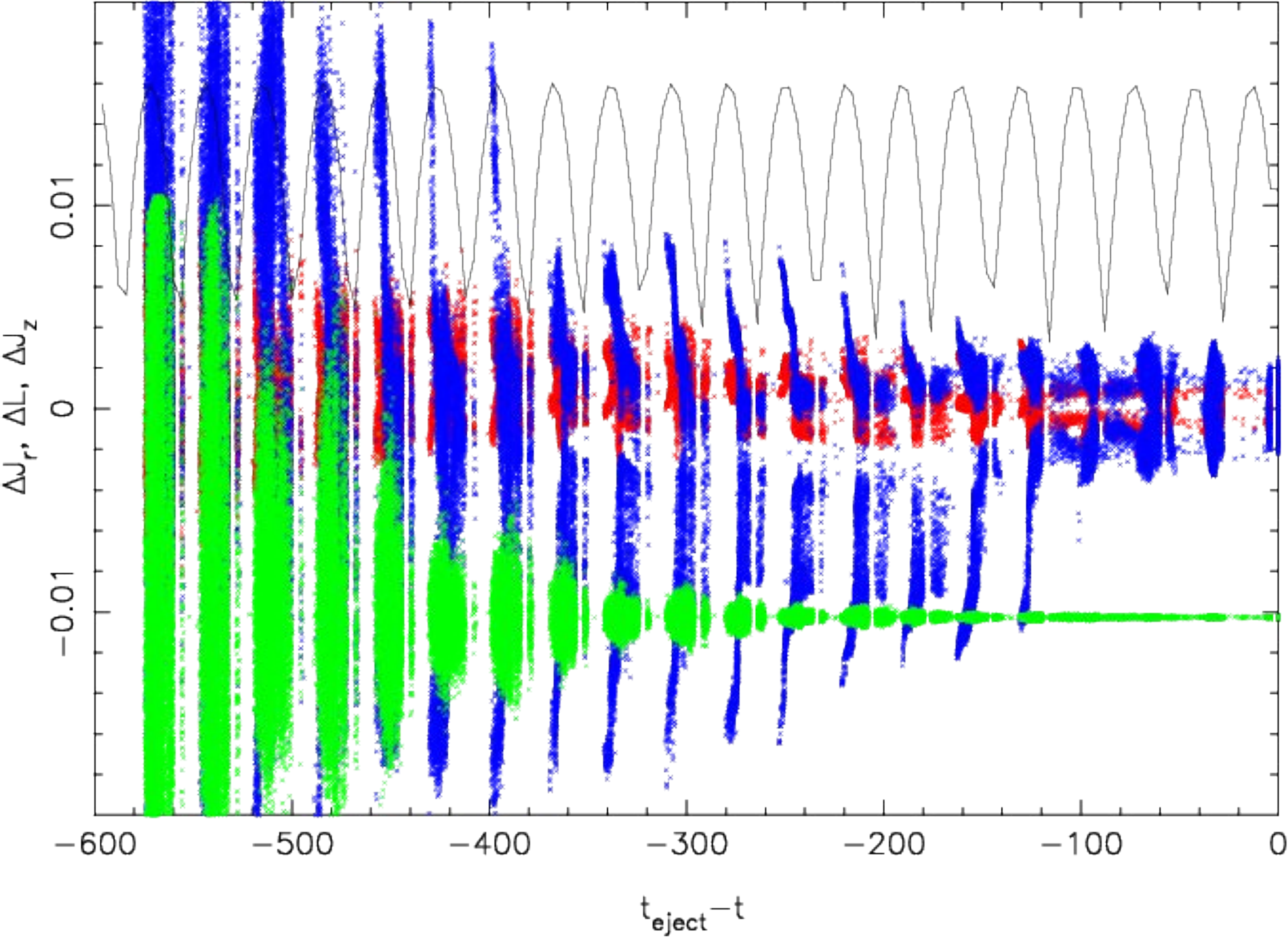} 
\end{center}
\caption{Same as Figure~\ref{fig_jrpz7h} but for satellites initiated on $L/L_c=0.4$ orbits. Note that the
vertical scale is a factor of two larger. The cluster essentially dissolves by the end of the simulation so the
tidal radius is reduced for more recently lost star particles.}
\label{fig_jrpz4h}
\end{figure}

An important issue is how dark matter sub-halos 
affect stream actions. 
The effects of a modest sub-halo fraction is
shown in Figures~\ref{fig_jrpz7h} and \ref{fig_jrpz4h} 
where 0.5\% of the halo mass of the same model potentials as in 
Figures~\ref{fig_jrpz7} and \ref{fig_jrpz7}
is converted to 1000 equal mass sub-halos.
The addition of sub-halos largely leaves the local character of the action distribution 
in place, a consequence of the probability of a sub-halo hitting the stream just as it emerges 
being fairly low. However, with increasing distance down the stream the cumulative action of the sub-halos
is to create sharp offsets in the actions that can be up to a factor of two changes relative to the satellite values.
However, the absolute size of the
changes are typically only 1\% of the underlying value so within a variation of about 1\%
the action variables along the stream can be considered approximately constant.

\section{ Low Eccentricity Stream Kinematics}

\begin{figure}
\begin{center}
\includegraphics[angle=-90, scale=0.4, clip=true, trim=45 30 30 40]{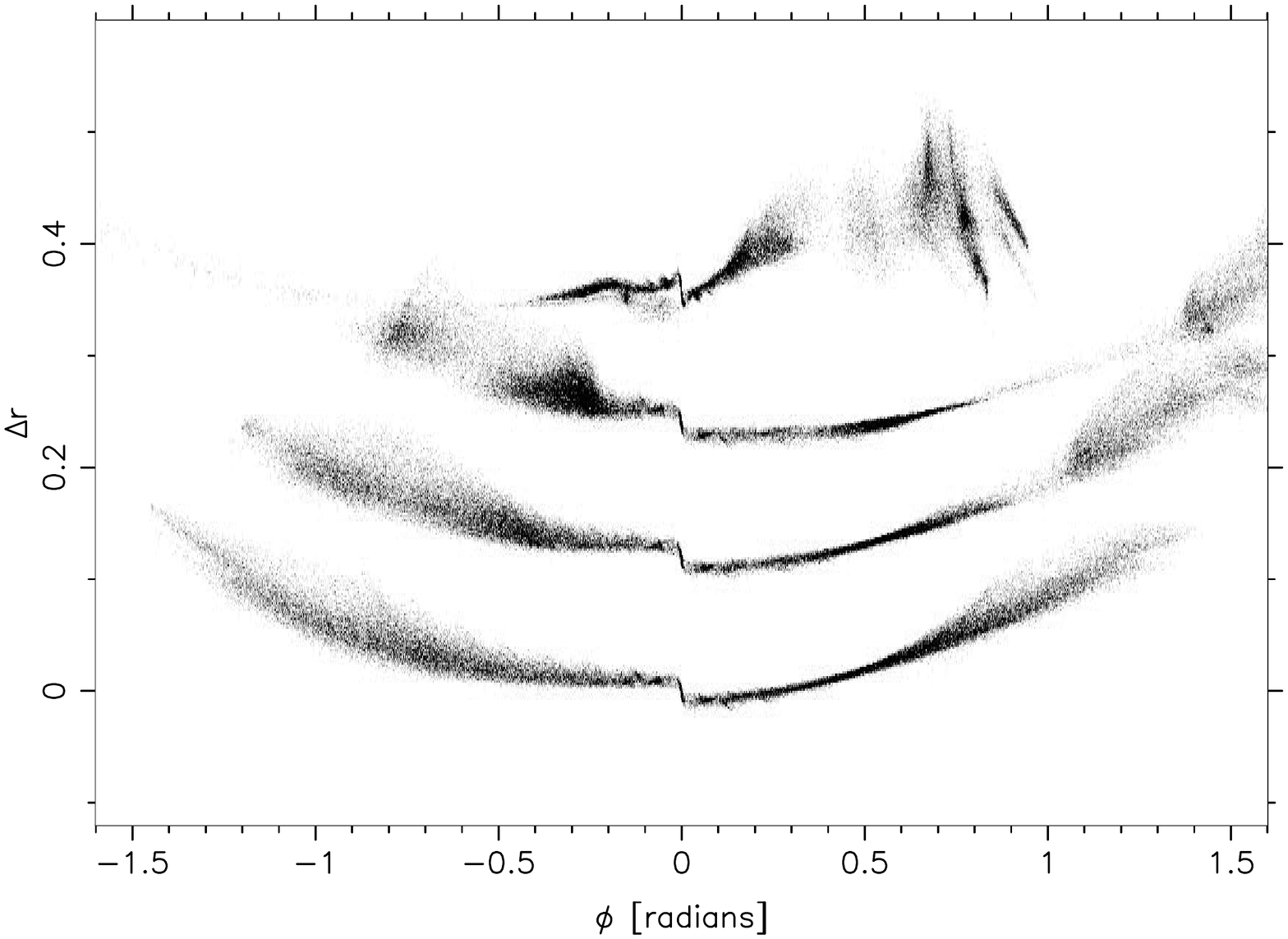} 
\includegraphics[angle=-90, scale=0.4, clip=true, trim=45 30 30 40]{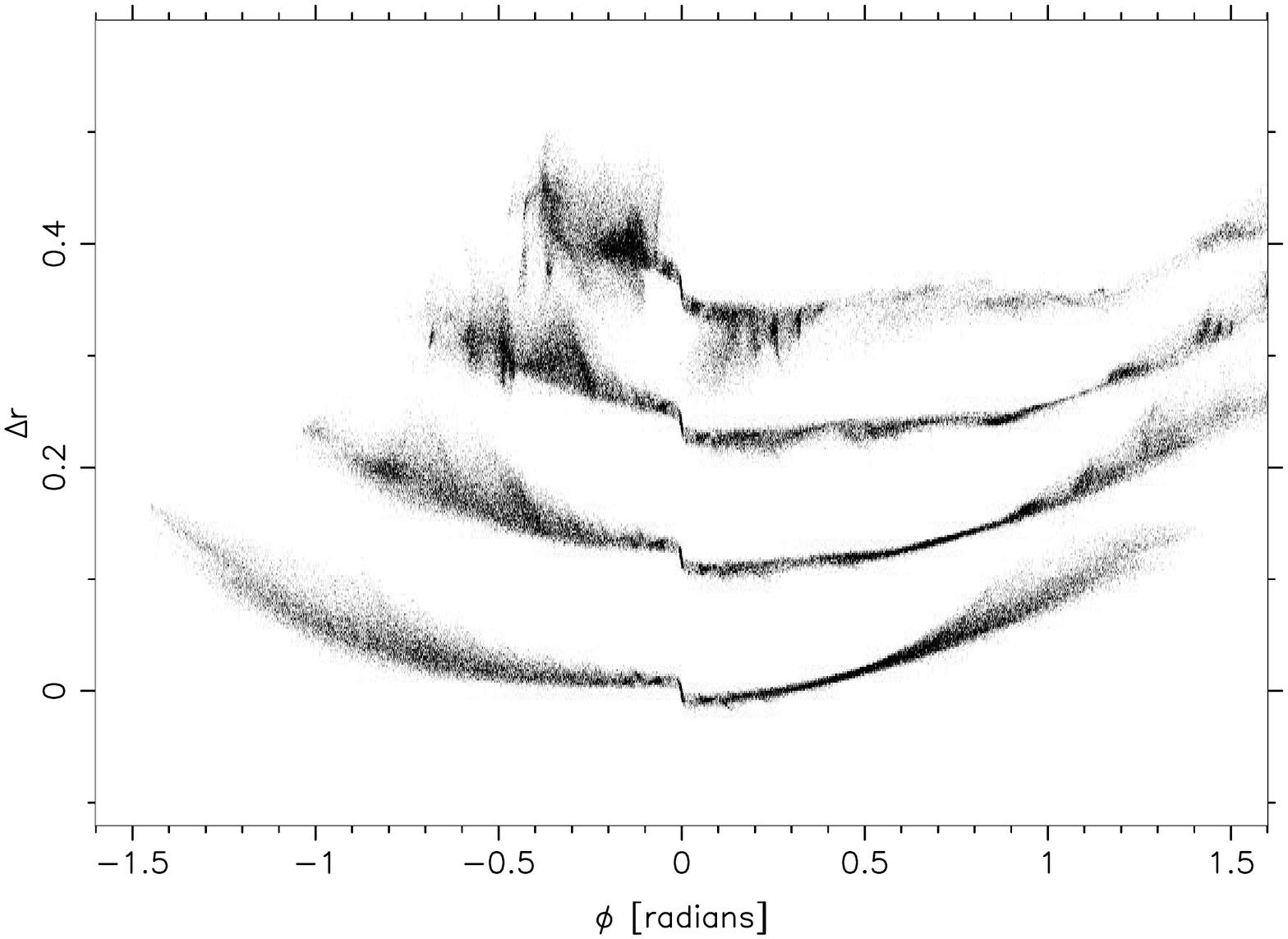} 
\includegraphics[angle=-90, scale=0.4, clip=true, trim=45 30 30 40]{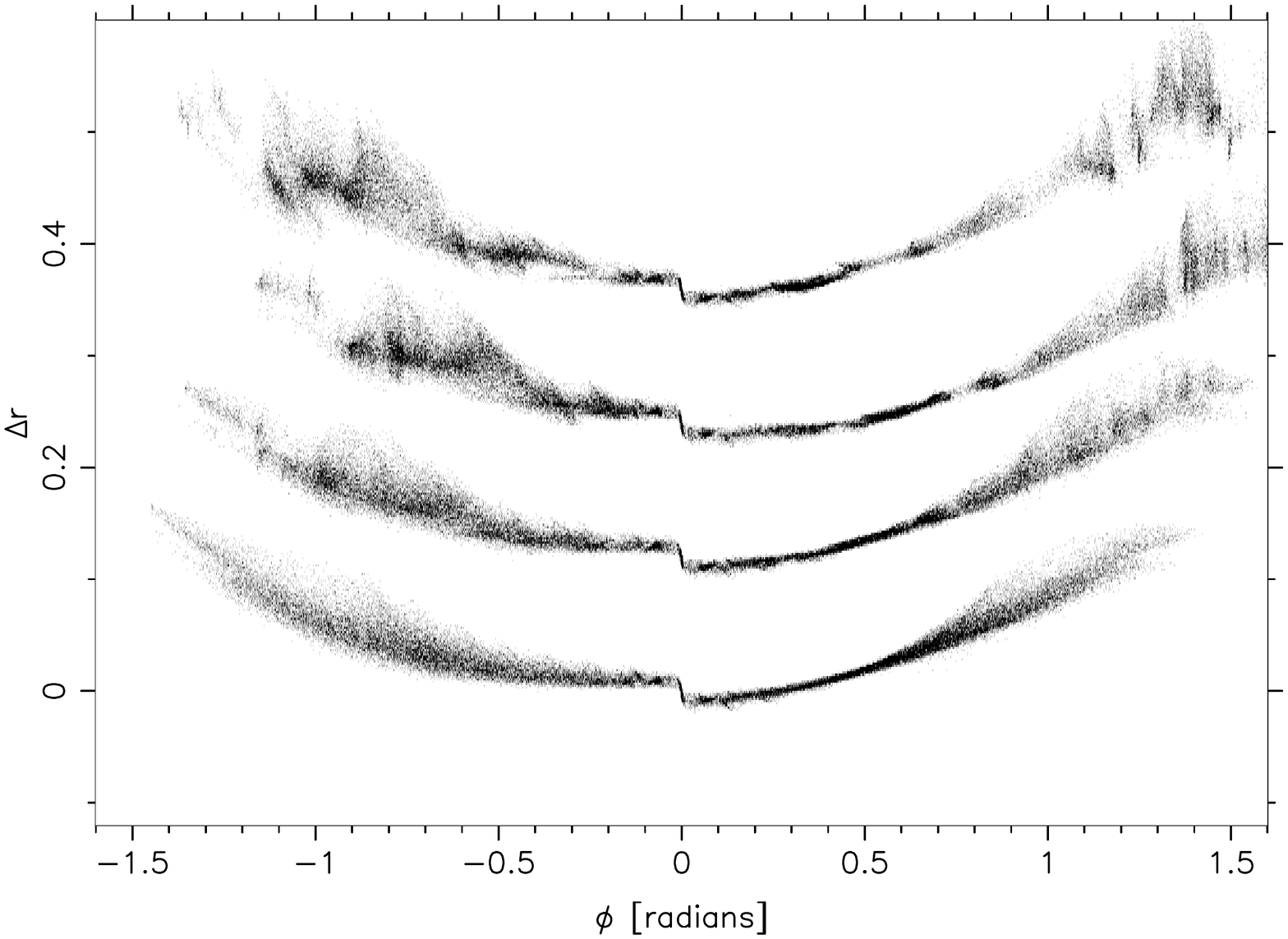} 
\includegraphics[angle=-90, scale=0.4, clip=true, trim=45 30 30 40]{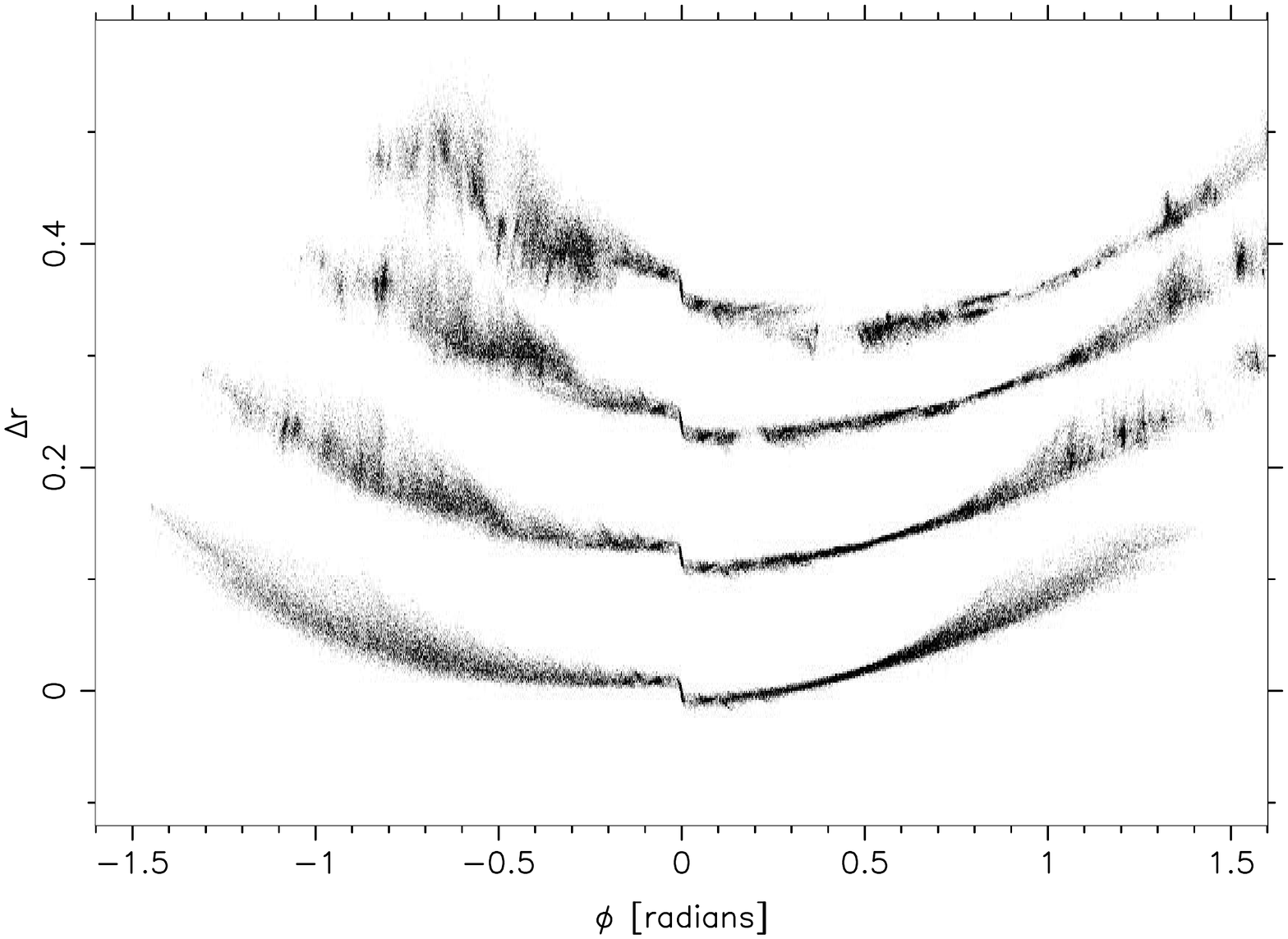} 
\end{center}
\caption{The stream locations relative to radius of the orbit of the satellite, $\Delta r$
 relative to the angular location of the satellite, $\phi$. The particles from
satellites on  $L/L_c=0.7$ orbits
with 100, 300, 1000 and 3000 sub-halos. 
The mass in subhalos in each panel is 0 at the bottom then 0.2\%, 0.5\%, and 1\% (top).}
\label{fig_drt7}
\end{figure}

\begin{figure}
\begin{center}
\includegraphics[angle=-90, scale=0.65, clip=true, trim=45 30 30 40]{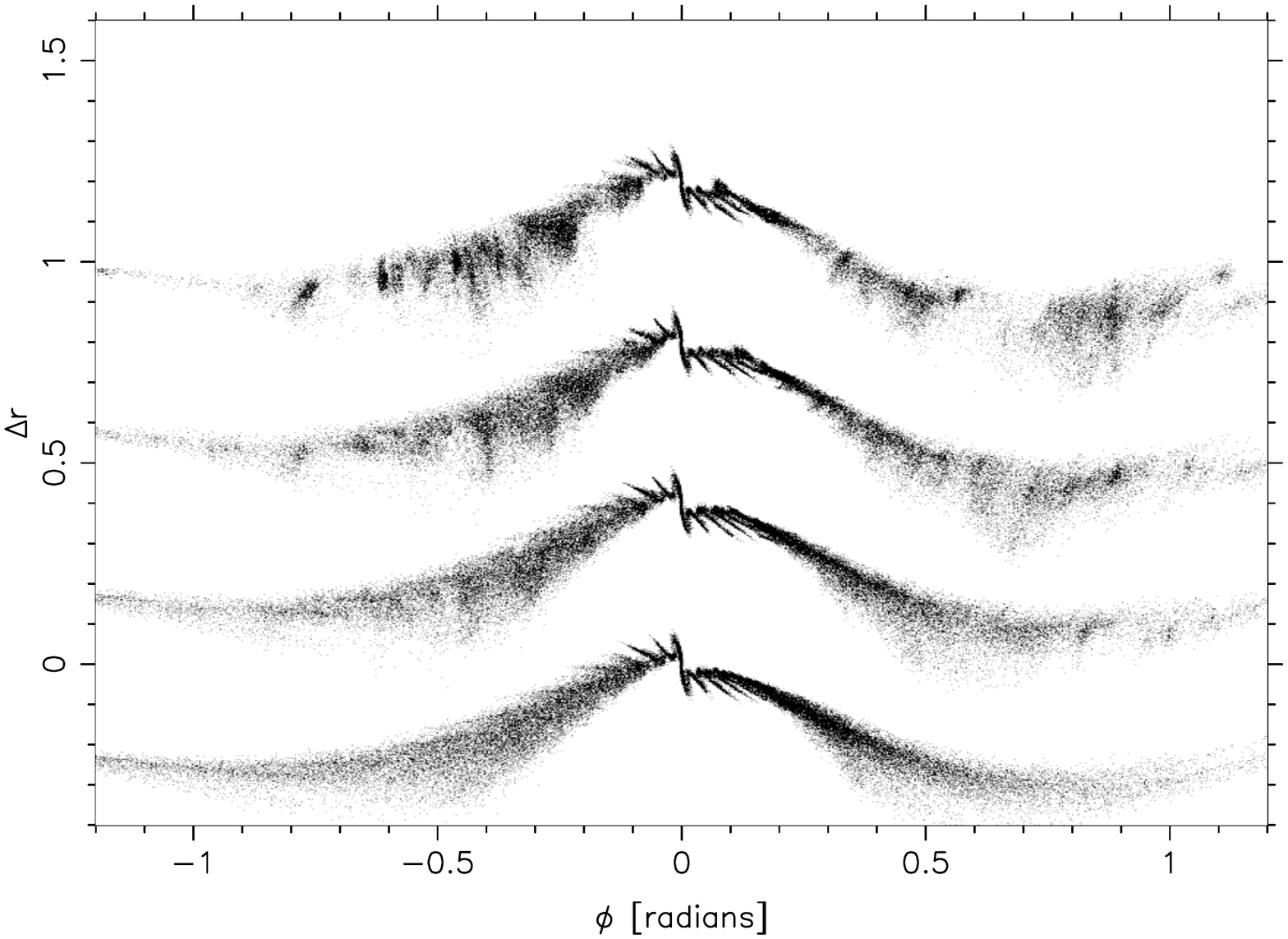} 
\includegraphics[angle=-90, scale=0.65, clip=true, trim=45 30 30 40]{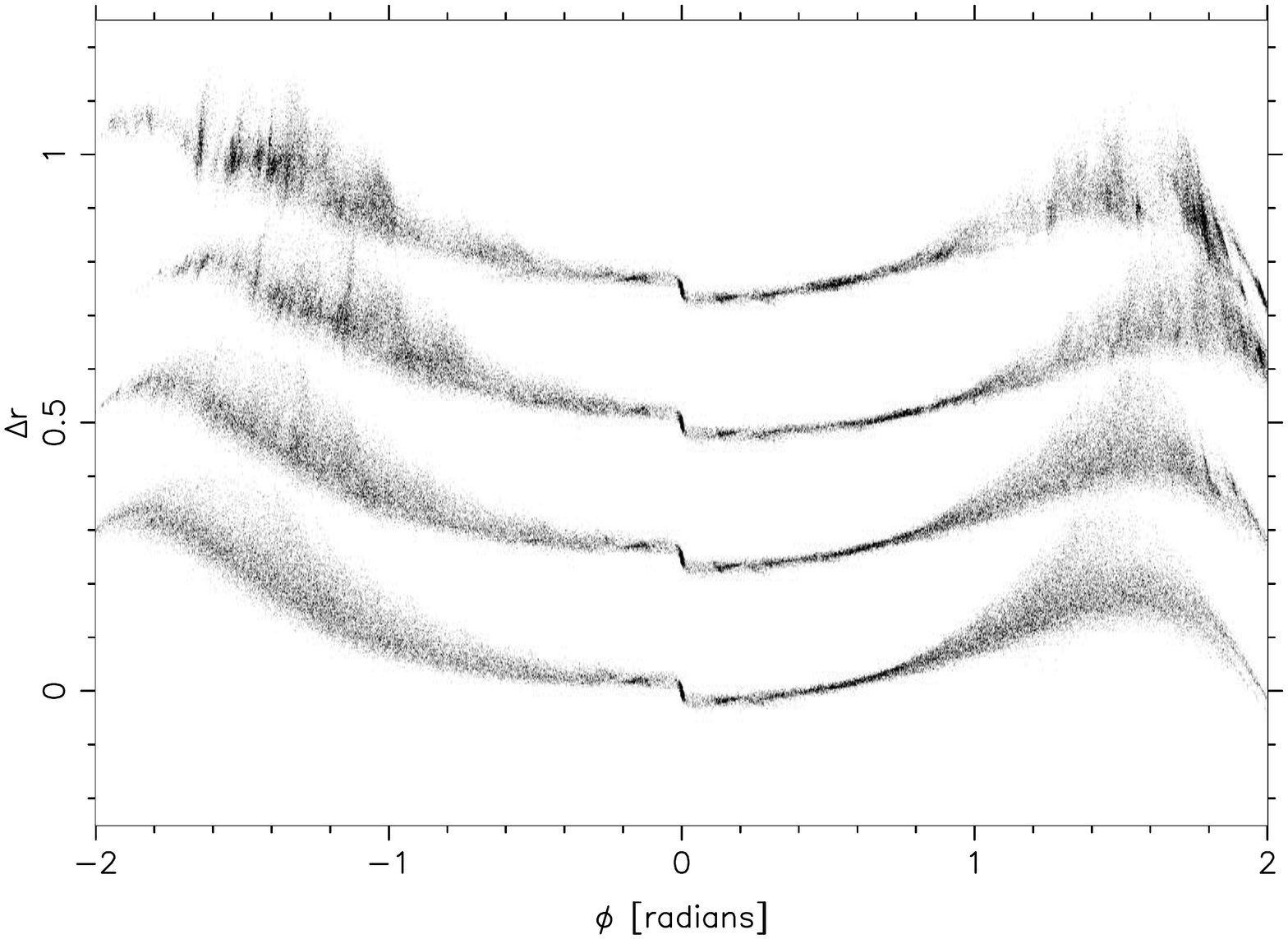} 
\end{center}
\caption{The $\Delta r-\phi$ plot for a satellite ten times
as massive as in Figure~\ref{fig_drt7}, $m=10^{-7}$ with 
no sub-halos at the bottom, then 1000 sub-halos 
with 0.2\%, 0.5\%, and 1\% (top)
of the halo mass, for time 580 (top) and 600 (bottom) to illustrate the dependence on orbital phase with a time
near apocenter at the top and near pericenter at the bottom. 
The bottom panel is comparable to the second from the bottom panel of Figure~\ref{fig_drt7} although the scales are 
different.}
\label{fig_drt7m}
\end{figure}

\begin{figure}
\begin{center}
\includegraphics[angle=-90, scale=0.4, clip=true, trim=45 30 30 40]{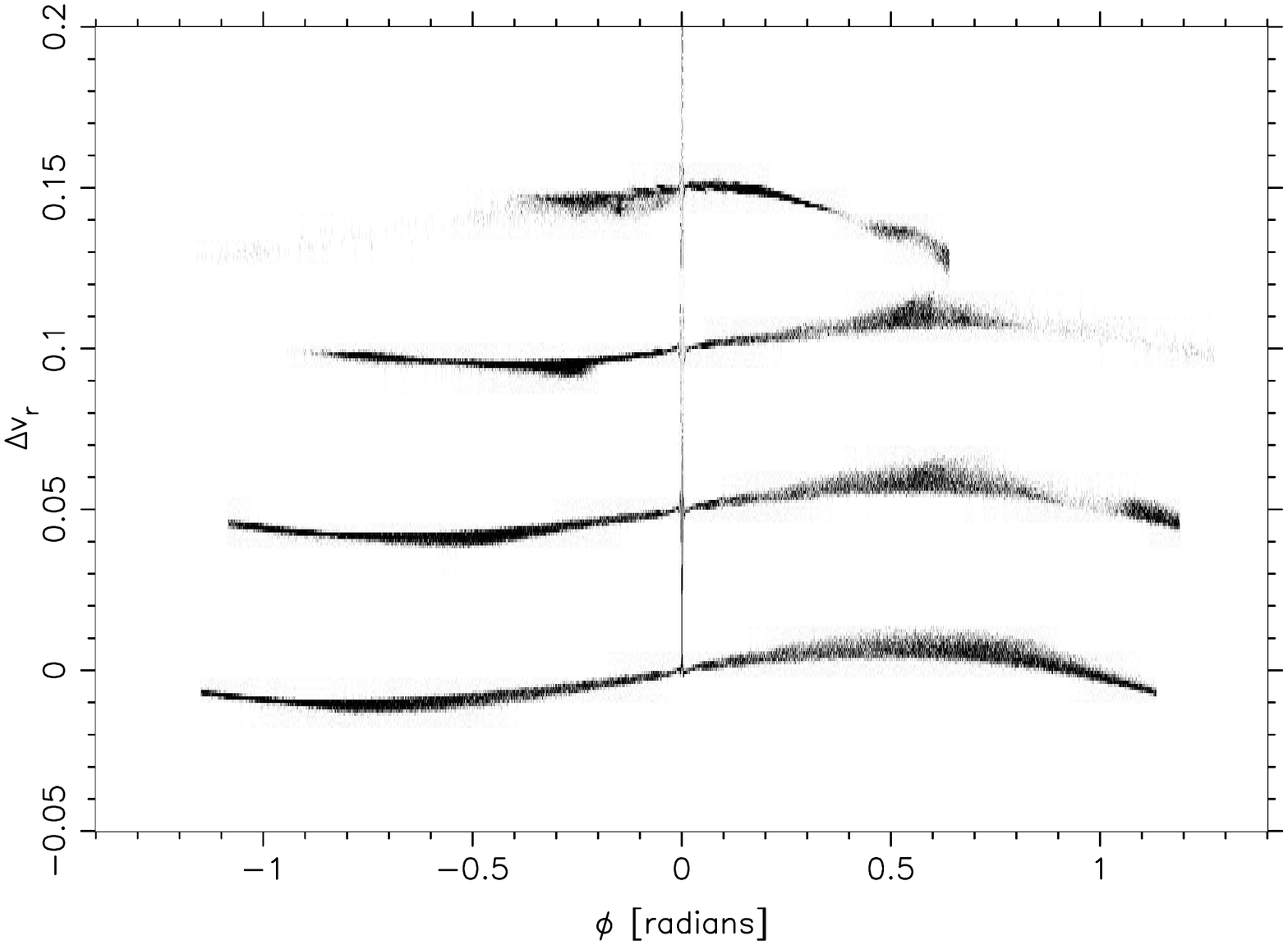} 
\includegraphics[angle=-90, scale=0.4, clip=true, trim=45 30 30 40]{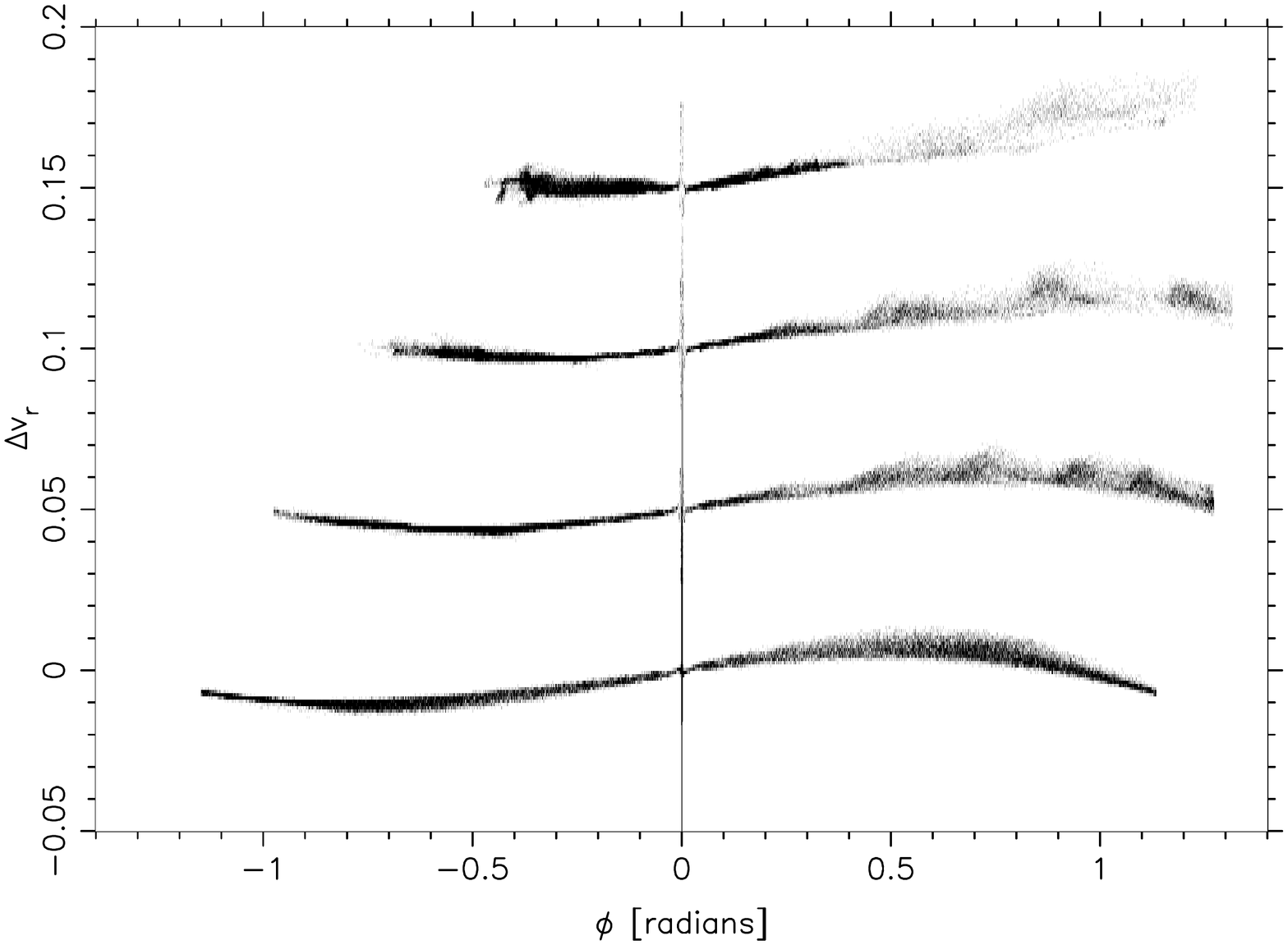} 
\includegraphics[angle=-90, scale=0.4, clip=true, trim=45 30 30 40]{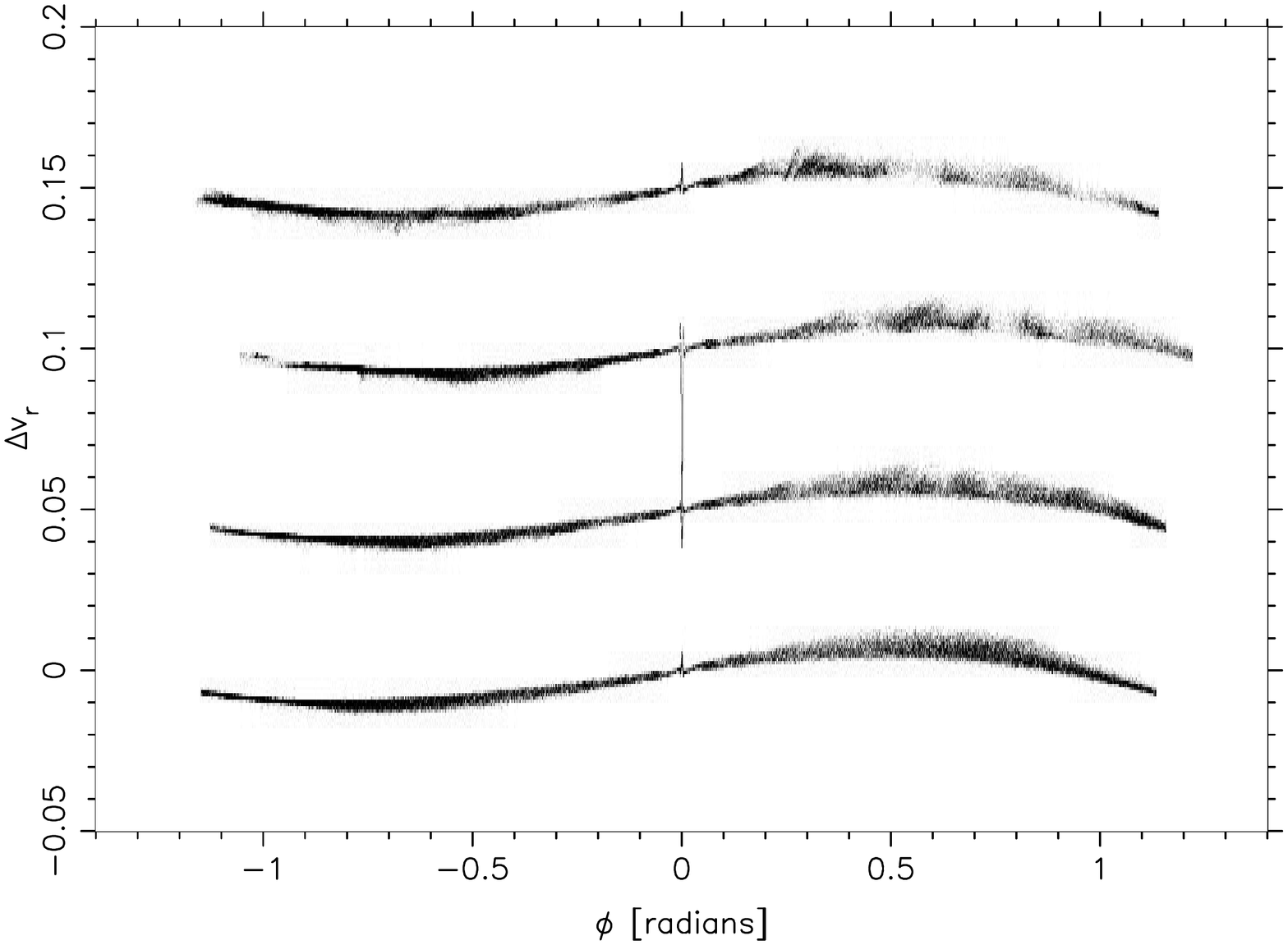} 
\includegraphics[angle=-90, scale=0.4, clip=true, trim=45 30 30 40]{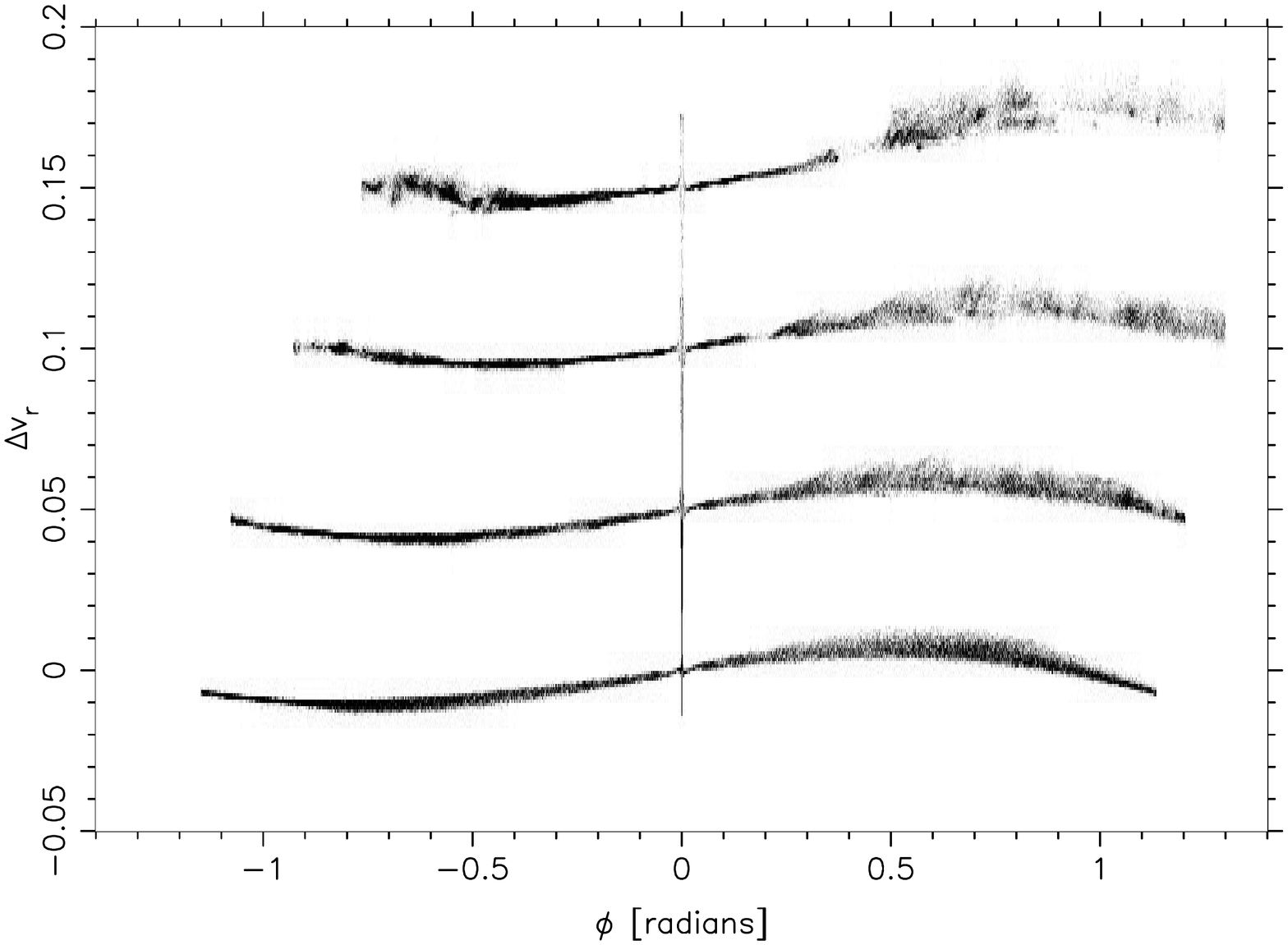} 
\end{center}
\caption{Same points as Figure~\ref{fig_drt7} except with the radial velocities relative
to the orbital radial velocity at that angle.}
\label{fig_vrt7}
\end{figure}

\begin{figure}
\begin{center}
\includegraphics[angle=-90, scale=0.4, clip=true, trim=45 30 30 40]{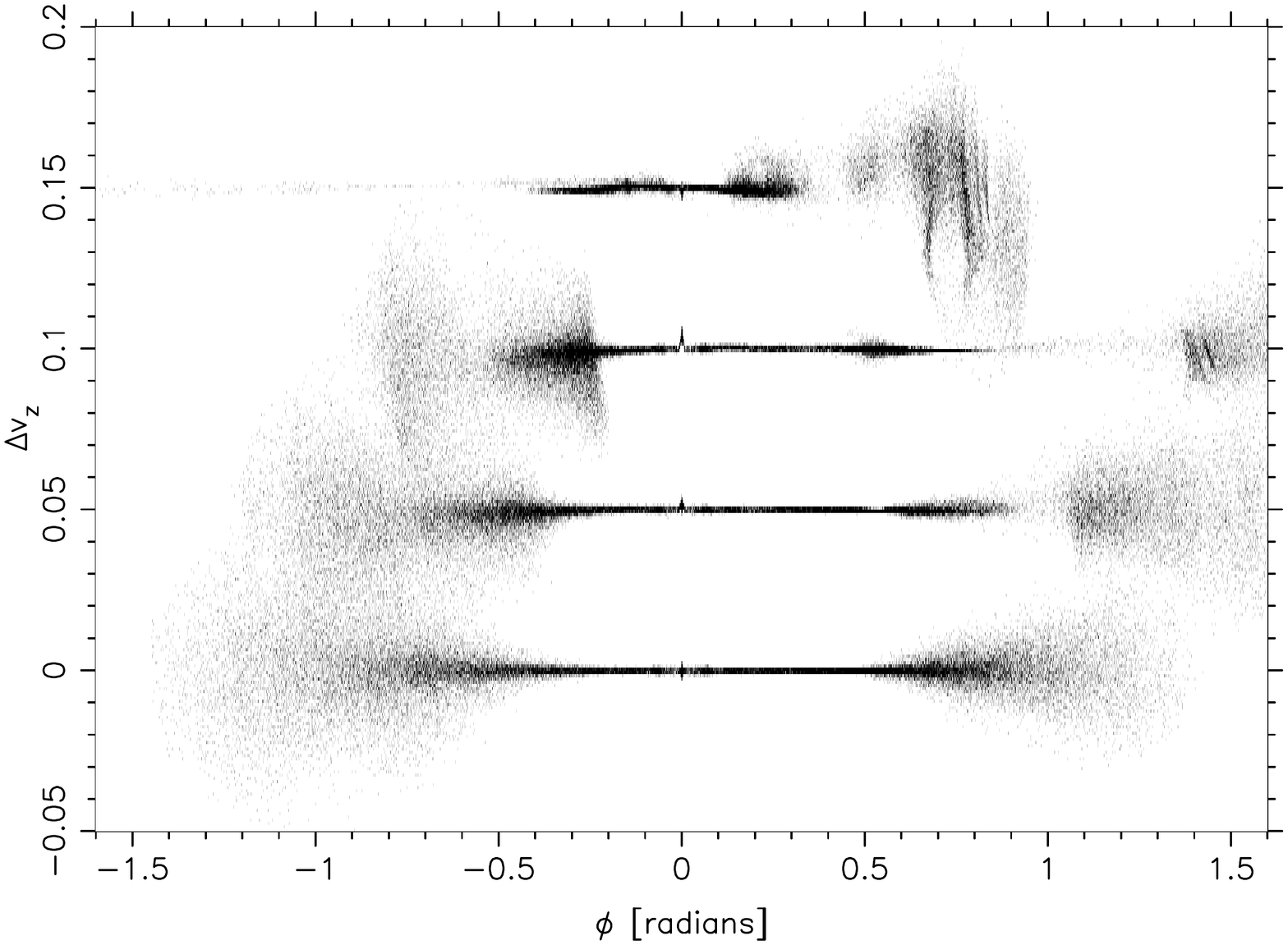} 
\includegraphics[angle=-90, scale=0.4, clip=true, trim=45 30 30 40]{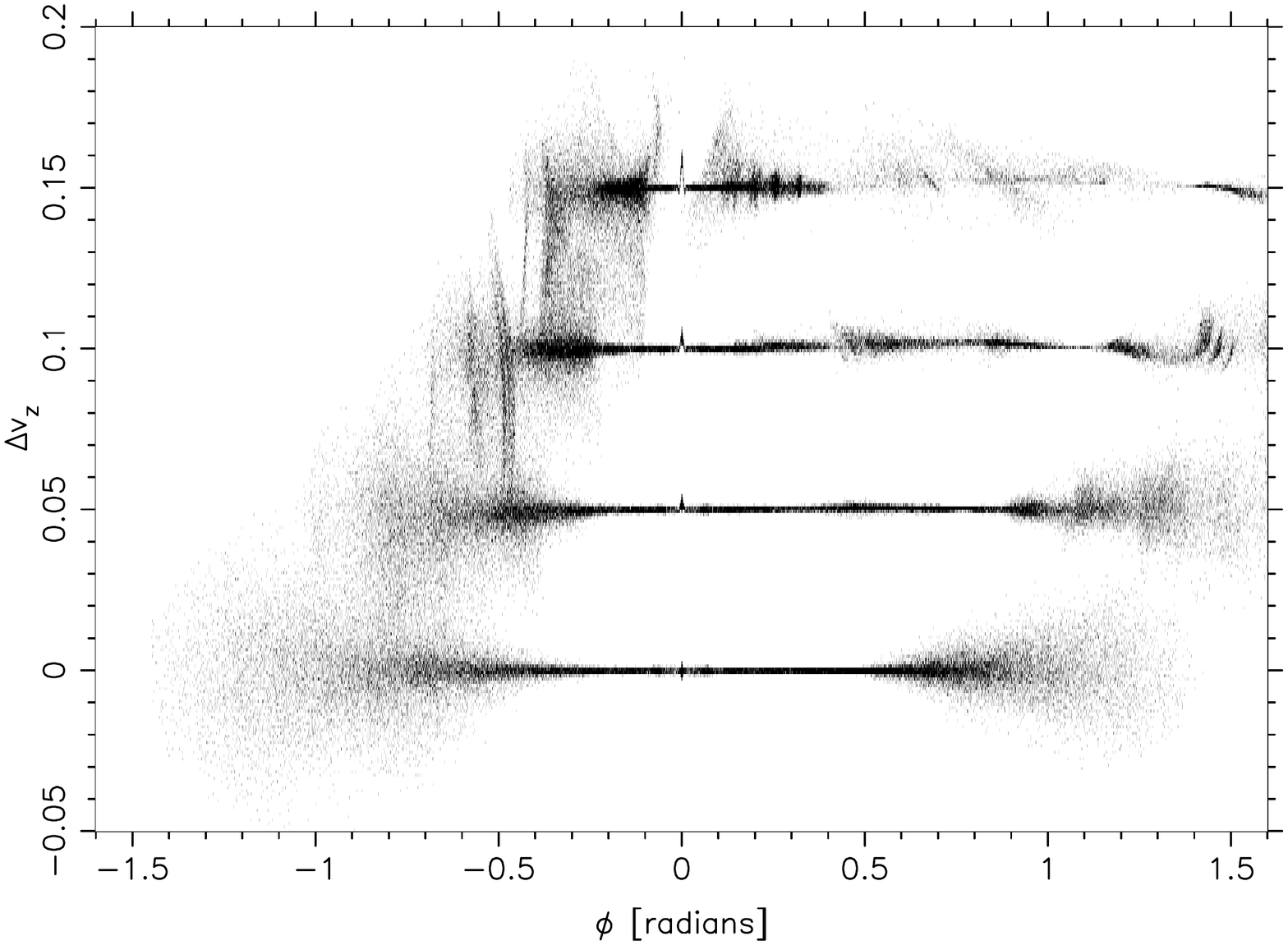} 
\includegraphics[angle=-90, scale=0.4, clip=true, trim=45 30 30 40]{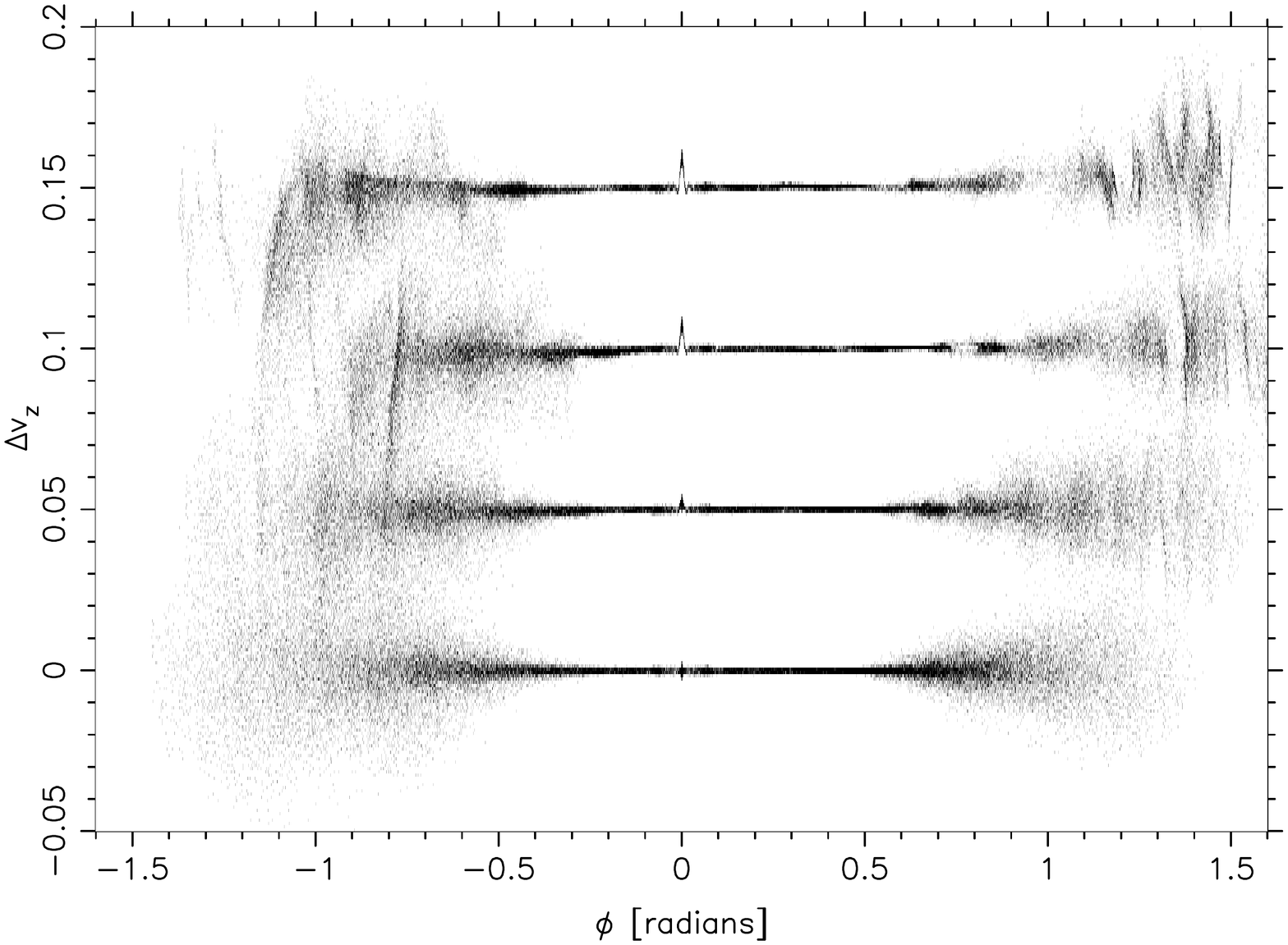} 
\includegraphics[angle=-90, scale=0.4, clip=true, trim=45 30 30 40]{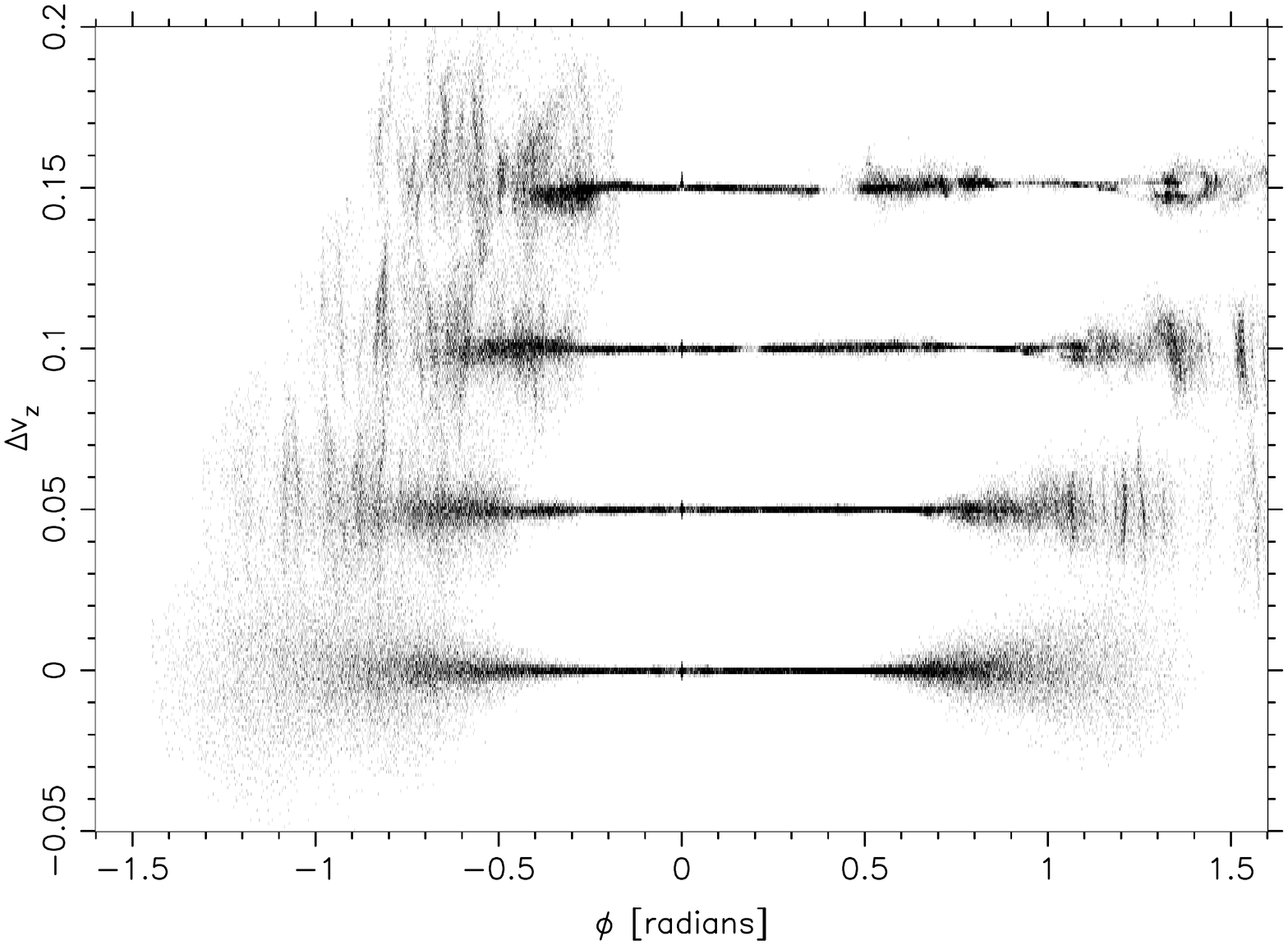} 
\end{center}
\caption{Same points as Figure~\ref{fig_drt7} except with the vertical velocities relative
to the orbital vertical velocity at that angle.}
\label{fig_vzt7}
\end{figure}

\begin{figure}
\begin{center}
\includegraphics[angle=-90, scale=0.7, clip=true, trim=45 30 20 40]{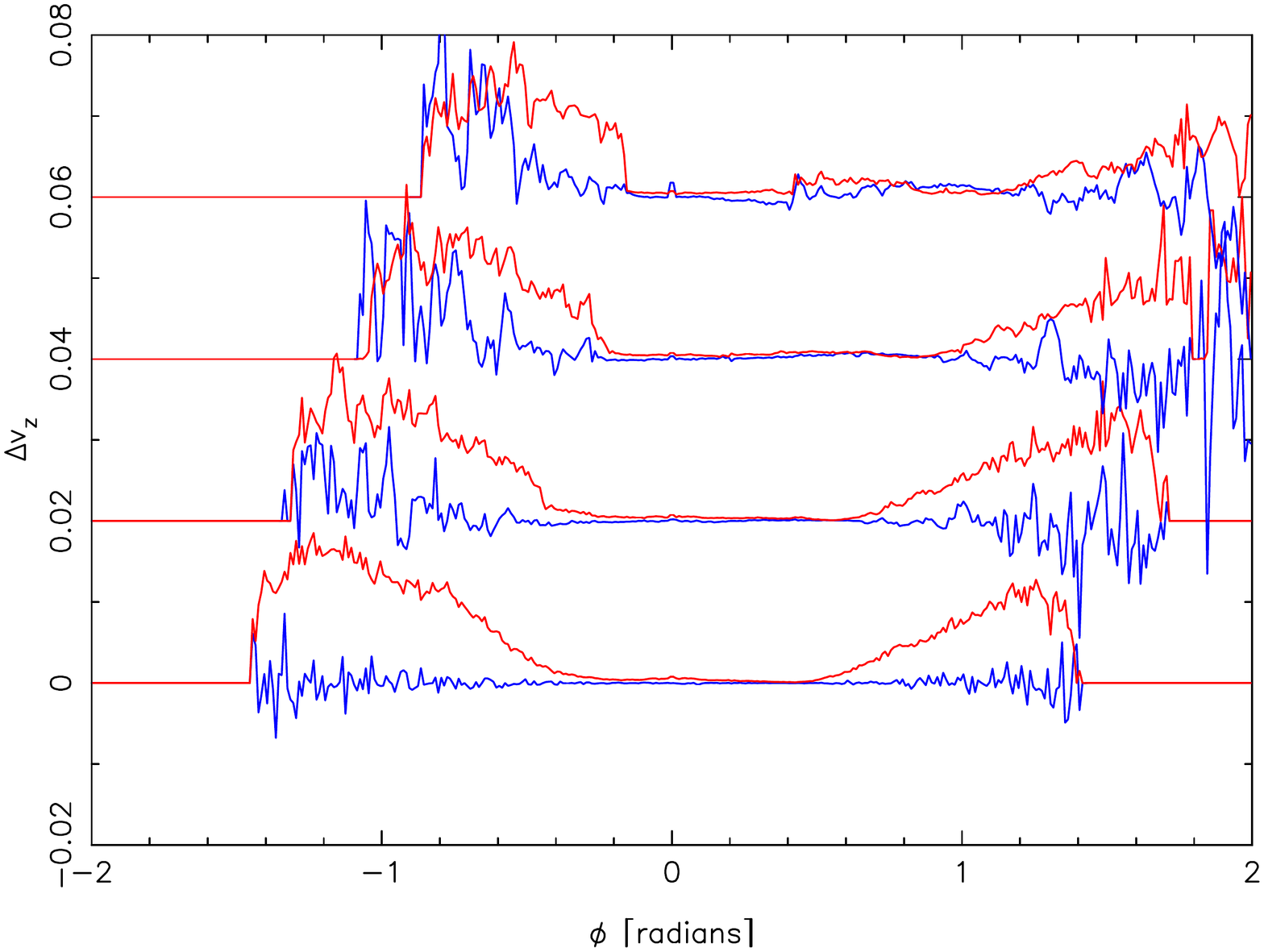} 
\end{center}
\caption{The mean (blue) and dispersion (red) in vertical velocities along the stream 
of Figure~\ref{fig_vzt7} for 1000 sub-halos, with the sub-halo mass
fraction rising from 0 at the bottom to 0.2, 0.5 and 1\% at the top.}
\label{fig_msv7f}
\end{figure}

Low eccentricity streams have less differential shear and less morphological change with orbital phase than 
high eccentricity streams \citep{Carlberg:15} making them better suited than high eccentricity
streams as indicators of the sub-halo population of the galactic halo.
Tidal streams started with $L/L_c=0.7$ in an $a_y=-0.1$, $a_z=-0.05$ halo with an 
increasing mass fraction in a varying number of sub-halos, ranging from 100 to 3000, 
along with zero for reference, are shown in Figures~\ref{fig_drt7}.
The plots show the radial position of the particles relative to radius of  the orbit of the center
of the star cluster at that azimuthal angle. 
The reference orbit is calculated  in the spherical isochrone potential at the end of the simulation.
The figures show that the sub-halos are sufficiently numerous and massive to have significant
encounters with the streams. At a 1\% sub-halo mass
fraction the stream is becoming so badly contorted for any of the sub-halo numbers 
that 1\% becomes an approximate upper limit to
the sub-halo fraction that can be tolerated for thin  streams, although this is consistent with the predicted local fraction.
 With the mass entirely in 100 or 300 sub-halos the stream encounters are rarer, but have
a larger effect leading to large gaps.
As the sub-halos become more numerous and less massive the stream
 is able to large retain its basic shape, with the
sub-halos producing gaps in the stream, as is most clearly seen for the 3000 sub-halo case.

The 1000 sub-halo case is repeated for a star cluster of mass $m=10^{-7}$ in Figure~\ref{fig_drt7m} which
is ten times the cluster mass of Figure~\ref{fig_drt7}. The stream width and length
scale with the tidal radius so are a factor of 2.15 times larger.  
Two snapshots are shown to illustrate the variation with orbital phase. 
The lower panel of  Figure~\ref{fig_drt7m} can be compared
to the second panel from the bottom of Figure~\ref{fig_drt7}, but note that the axes are not the same.
Although the sub-halo effects remain constant, the figure also shows that sub-halo induced gaps 
remain visible in this slightly more than twice as wide stream from a ten times more massive progenitor cluster.
Figure~\ref{fig_drt7m} also shows that the tidal ``spurs'' on the stream are much more evident particularly
near apocenter as a consequence of the orbital phase dependence of the mass ejection from the satellite being initiated
at pericenter with the particles drifting away at apocenter as is the case for Pal~5 \citep{Dehnen:04}.

The radial and vertical velocities of the 
streams emanating from the  $m=10^{-8}$ cluster
of Figure~\ref{fig_drt7} are shown in Figures~\ref{fig_vrt7} and \ref{fig_vzt7}.  
The effects of increasing mass fraction in sub-halos, $f_{sh}$,  over a range of $N_{sh}$ are clearly visible
in the velocities of \ref{fig_vrt7} and \ref{fig_vzt7}.
In all cases the amplitude of the perturbations grows significantly with increasing $f_{sh}$ which
gives a better defined ``gap'' along the stream.
The radial velocities mix together the velocities of the tidal features and have considerable orbital 
phase dependence, whereas the velocities perpendicular to the orbital plane have little
orbital phase dependence. 
The mean vertical velocity and its dispersion about the mean are shown in Figure~\ref{fig_msv7f}.
Along the stream the dispersion rises as the orbits spread out in the aspherical potential 
\citep{HW:99,SGV:08,Ngan:15b}. 
Further down
the stream the stream has been out of the cluster for a longer time and has been subject to more
sub-halo strikes. The mean velocity is knocked away from zero and there are local increases 
in the velocity dispersion. on a length scale comparable to the scale size of
the dark matter halo. The Hernquist radius of the sub-halos
is $0.017 \sqrt[3]{(f_{sh}/0.005)(1000/N_{sh})}$. For
an orbital apocenter of 30 kpc, the sub-halo radius scales to 0.13 kpc.
Kinematic 
detection of sub-halo crossings requires enough velocity measurements per kiloparsec
to be able to detect the expected velocity changes. The numbers can be worked out from simulations
matched to a stream of interest, but it will be of order one hundred measurements of stream velocity per kiloparsec
to be able to get started. 
For 0.5\% of the halo mass in sub-halos, excursions around the mean velocities  reach 0.02 units which is 5\% of the  circular velocity of 0.4 units.
Or, if scaled to a circular velocity of 220 \kms\ means that values of 11 \kms\ are expected in the largest
 excursions from the mean. 
For 0.2\% mass in sub-halos the velocity excursions are proportionally smaller.
The velocity dispersion in
the thin parts of the streams, which are most readily identified, is about 0.002 units, which scales to about 1\kms, so these
excursions should be readily detectable given enough velocities along the stream. 

\section{Discussion and Conclusions}

We have developed a framework which is able to reproduce a large range of properties of tidal streams 
as are seen in galactic halos extracted from cosmological simulations.
Basing the potential 
on the isochrone puts these simulations into a well studied
regime and allows easy dynamical insight into the orbital mechanics at play.
A central element of tidal stream analysis 
 is that the distribution of star velocities in the stream can be 
accurately calculated, here with a spherical shell n-body code. 
Once free of
the satellite potential,  stars orbit freely, 
responding to in the overall potential around the orbit
and the presence of sub-halos orbiting within  the host potential. 
A question of interest is how much do the orbital actions of stream stars change relative to their initial
values as a result of the asphericity of the potential and any sub-halos. If the changes are large it would 
diminish the value of streams as probes of the potential shape.
We find that for realistic local sub-halo mass fractions of 1\% or less that the orbital actions changes
are about 1-2\% over approximately 15-20 radial orbits. The  stars most distant from the progenitor cluster
are the most affected, but the changes are small enough that streams can be approximated, to this precision, as having invariant 
orbital actions.  The outcome is a desirable situation in which sub-halos should have easily measurable local effects in velocity 
space, but the mean stream motions are not greatly disturbed.

Tidal streams in spherical halo models are much thinner and simpler than
 those seen in more realistic cosmological halos, whether there are sub-halos or not \citep{Ngan:15b}.
The underlying cause is that aspherical potentials have orbits that are not present in a spherical potential, opening
up  more ways for the stream to spread.  Near the progenitor the stream is sufficiently young that the number of 
sub-halo encounters is low and the stream dynamics mainly reflect the change in tidal field around the orbit.
Further down the stream, where the stream has been orbiting freely longer, the cumulative effect of sub-halos and potential asphericity builds up.
In the case of a high ellipticity, $e\gtrsim 0.6$, star cluster progenitor, the stream orbit spreading
can be so large 
that the majority of the stream would be hard to identify as a high density region on they sky.
However, the dense segments of high eccentricity streams certainly are visible, even though they may
represent only a few percent of the entire stream, leaving the rest as a ghost. 
It is therefore possible that a short stream segment like Ophiuchus could be the high density region of a stream with
a total luminosity a factor of ten or more larger. 

Velocity measurements are the key to making progress with stream dynamics as indicators of both the shape of the halo and
the substructure in the halo. It is now straightforward to reliably predict the velocity 
distribution of the particles newly injected into the stream, 
provided that the mass loss is predominantly driven by galactic tides.  
The presence of sub-halos
leads to patches along a stream comparable in size to the 
characteristic radius of a sub-halo, typically 0.1-0.3 kpc, with an enhanced velocity dispersion and velocity
offsets from the mean value for the stream, as shown in Figure~\ref{fig_msv7f}. 
It will likely be easier to detect the mean velocity offsets along the stream, rather than local velocity dispersion increases.
 The large scale shape of the galactic potential leads to smooth variations along the stream. 
Although there already is considerable incentive to acquire spectroscopy of stars in stellar streams 
for abundances and orbital analysis, 
 these simulations 
show that once large numbers of precision velocities are available for stream stars 
they will provide definitive tests for the presence of sub-halos.

This paper has concentrated on how sub-halos and triaxial potentials effect action variables, orbits and velocities and
left aside an examination of gaps in streams \citep{Yoon:11,Carlberg:12,CG:13,Erkal:15} 
as indicators of the sub-halo population. Gaps will be most useful as indicators in low eccentricity streams, 
being blurred out fairly quickly in high eccentricity streams \citep{Carlberg:15}. 
An analysis 
of gaps and velocity offsets in these streams as sub-halo indicators will be presented elsewhere.

\acknowledgements

This research was supported by CIFAR and NSERC Canada.

\end{document}